# Comparative study of equilibrium and non-equilibrium predictions by different models for a hypersonic cone at high-altitude

Mengyu Wang[a], Pan Yan[a], Qin Li[a*], Zhenfeng Wang[b*], Xiaoming Guo[b], Yuanchun Liu[b]

a. School of Aerospace Engineering, Xiamen University, China, 361102

b. Science and Technology on Space Physics Laboratory, China Academy of Launch Vehicle Technology, Beijing 100076, China

**Abstract:** As one of fundamental configurations of hypersonic vehicles, the cone as well as its aerodynamic characteristics and gas physics under extreme flight conditions are always focused by engineering. Targeting a cone with the half-angle as 10° at $M_\infty$ = 27 and $H$ = 72 km, simulations were conducted comparatively to analyze the predictions by different models which not only include the equilibrium and diverse thermochemical non-equilibrium gas models, but also involve slip and non-slip boundary conditions. Following validation and grid convergence studies, the computational methods are verified and the grid scale is determined. Then, systematic comparisons on aerodynamic performance, flow structures, and characteristic distributions were performed to understand the equilibrium and non-equilibrium characteristics as well as their discrepancies. The key findings are: (1) While the overall flow structures are broadly similar, discrepancies exist in the features at the base locations, e.g., the diverse high-temperature distributions. Notably, the vibrational temperatures distribute differently under slip and non-slip boundary conditions near the wall; (2) The equilibrium gas model predicts higher drag coefficient, wall heat flux, and skin friction than those of non-equilibrium models. Predictions also vary among the non-equilibrium models themselves. Specifically, compared to the three-temperature model, the one- and two-temperature models predict larger drag coefficients with the relative difference exceeding 5%. Nevertheless, the results from the three-temperature model with and without slip conditions are largely consistent; (3) The disparities between equilibrium and non-equilibrium characteristics are primarily manifested in the shock layer and wake regions. Within these regions, the overall temperature for the equilibrium gas is lower than that for the non-equilibrium cases, while in the latter specific non-equilibrium features are distinctly exhibited, e.g., the translational-rotational temperature is generally higher than that from the one-temperature model, and the vibrational-electronic temperature shows the opposite trend. Notably, in the slip flow within the three-temperature model, the translational-rotational temperature is higher and, particularly, the vibrational temperature is even larger than counterparts of the non-slip flows near the wall and base center line. In short, comprehensive understanding is achieved on equilibrium and non-equilibrium characteristics, which would yield valuable insights for engineering.

## 1 Introduction

Slender cones and similar geometries are fundamental configurations for typical hypersonic vehicles. Under high-altitude hypersonic conditions, severe air compression and viscous effects within the shock layer [1] create an extreme thermal environment for them. This triggers complex physicochemical processes, including gas dissociation and the redistribution of internal energy into vibrational and electronic modes [2]. These complexities render the ideal gas model inadequate. To accurately predict key aerodynamic and thermal properties, numerical simulations must employ models that capture real-gas effects, such as equilibrium and thermochemical non-equilibrium models. A detailed analysis of the prediction discrepancies between these models offers valuable insights for engineering applications. However, due to the high fidelity required of the underlying physicochemical models, researches in this area remain relatively limited.

---

[*] Qin Li and Zhenfeng Wang are co-corresponding authors, email: qin-li@vip.tom.com

Regarding comparative studies on the predictions of cone flow characteristics by different models, Singh et al. [3] investigated the influence of the thermochemical equilibrium model. They performed predictions of wall heat flux and pressure for cones with different nose radii at $M_\infty$ = 10 and 20, using both ideal gas and equilibrium gas models (*EGM*). Their results indicated that the two models exhibited only minor differences at high Mach numbers (*Ma*). Zeng et al. [4] performed numerical simulations for the RAMC-II configuration at high-altitude, high-speed conditions ($H$ = 61 km, $M_\infty$ = 23.9 and $H$ = 81 km, $M_\infty$ = 28.3) using one- and two-temperature models (*1T*- and *2T*-M) along with a nonlinear coupled constitutive relation model. Their results demonstrated that the *2T*-M predicted higher translational-rotational temperatures ($T_{tr}$) along the stagnation line and higher wall friction compared to the *1T*-M. Additionally, the electron number density near the wall showed better agreement with experimental data, indicating the presence of significant vibrational non-equilibrium effects. Zhong et al. [5] employed the DSMC method to simulate a 10° half-angle cone at $H$ = 80 km and $M_\infty$ = 27.5. Their results showed that chemical reactions were weak throughout the flow field. Translational and rotational modes were near equilibrium, while vibrational non-equilibrium was present. A low-density, high-temperature region was observed in the center of the wake recirculation. Furthermore, temperature jump was evident at the base wall, though no velocity slip was detected. Notably, Clarey et al. [6] conducted numerical simulations for the RAMC-II configuration under two hypersonic conditions, i.e., $H$ = 61 km, $M_\infty$ = 24.5 and $H$ = 71 km, $M_\infty$ = 25.9, using *2T*- and *3T*-M (three-temperature model). A comparison revealed that the electron number densities predicted by the different temperature models in the forebody region were similar and both agreed well with experimental data. However, comparisons of temperature distributions in the forebody and wake regions showed that while the $T_{tr}$ and vibrational temperature ($T_v$) profiles were largely consistent between the two models, the electronic temperature ($T_e$) along the stagnation line in the *3T*-M was slightly lower than $T_v$ due to relaxation processes. Furthermore, a distinctly non-equilibrium electronic state and corresponding electronic energy were observed in the wake region.

In addition, research has also been conducted on the effect of angle of attack (AOA or $\alpha$), particularly focusing on that on the base flow. Qin et al. [7] conducted numerical simulations on cones at an altitude of $H$ = 24.38 km with varying $\alpha$ and *Ma*. Their study revealed that as the $\alpha$ increased, the shear at the cone base intensified while the recirculation zone diminished. Furthermore, the recirculation, the base stagnation point, and the peak heat flux all shifted toward the leeward side. Clarey et al. [6] further investigated the effect of $\alpha$ by employing *3T*-M in numerical simulations of a 6° half-angle cone under the conditions of $H$ = 40, 50 km and $M_\infty$ = 25.9. The results indicated that when the $\alpha$ exists, the excitation of vibrational and electronic modes develops on the leeward side and extends into the wake field. As the $\alpha$ increases, the high-temperature region of $T_v$ expands significantly, its peak value rises, and the electron number density increases markedly. Additionally, Satchell et al. [8] performed numerical simulations using *2T*-M for a 7° half-angle cone under the conditions of $H$ = 30 km, $M_\infty$ = 7 and $\alpha$ = 5°. Their results demonstrated that structures such as the boundary layer and viscous mixing region formed near the head persist along the cone body into the wake, influencing its flow structure and species distribution. When the freestream is at an $\alpha$, the windward side experiences greater heat conduction than the leeward side, and also features higher temperatures and concentrations of chemical reaction products [8, 9].

In summary, research demonstrates that the equilibrium and non-equilibrium characteristics of hypersonic cones provide a critical reference for developing vehicles with speeds alike. These findings also serve as a foundation for subsequent theoretical studies and engineering applications, such as the construction of similarity parameters [10], the development of transport models [6, 11], and the analysis of wall catalysis/ablation [8, 9], as well as communications blackout and plasma sheath phenomena [12, 13]. Concurrently, existing research also reveals the following insufficiencies: (1) While several studies have examined the influence of gas models on blunt-nosed vehicles (e.g., the RAMC), investigations specifically focusing on cone configurations remain relatively scarce, and the sharp nose of the latter may lead to shock structures and overall non-equilibrium characteristics distinct from those of the former; (2) There is a lack of comprehensive understanding regarding their equilibrium and non-equilibrium (particularly under the *3T*-M and more complex thermal non-equilibrium models) properties and the discrepancies among them. This is especially true for the characteristics within the continuum and near-continuum flow regimes under high-altitude, high-*Ma* conditions, with the properties of the wake flow field being particularly poorly understood. With the advancement of hypersonic vehicle technology in recent years, the base flow has garnered increasing attention and emerging engineering requirements place a strong emphasis on its flow structure and physicochemical characteristics [14, 15].

To address the issues outlined above, this study selects a 10° half-angle cone for numerical simulations under high-altitude ($H$ = 72 km), high-*Ma* ($M_\infty$ = 27) freestream conditions at $\alpha$ = 0° and 8°. Simulations are conducted using thermochemical equilibrium and non-equilibrium models. A systematic comparison is performed of the differences in flow structure, aerodynamic characteristics, and equilibrium/non-equilibrium properties predicted by various real-gas models, including the equilibrium gas model, the chemical non-equilibrium model, and multi-temperature thermochemical non-equilibrium models as well as slip/non-slip boundary conditions. Particular emphasis is placed on the analysis and discussion of the complex flow within the wake region. To enhance simulation accuracy and resolution, the numerical scheme employs the high precision and robust WENO3-$Z_{ES3}$ [16]. The paper is organized as follows. Section 2 introduces the governing equations, thermochemical equilibrium and non-equilibrium models, boundary conditions, the WENO3-$Z_{ES3}$ scheme, and computational validation cases. In Section 3, a grid convergence study is first conducted; subsequently, simulations of the cone under different models and boundary conditions are performed, comparing their predicted aerodynamic characteristics and flow structures. A qualitative and quantitative analysis of the equilibrium and non-equilibrium properties in the forebody and wake flows is then carried out. Finally, conclusions are provided in Section 4.

## 2 Numerical methods

2.1 Governing equations

Considering the high altitude of the vehicle under study, the simulations utilize the laminar Navier–Stokes equations combined with either equilibrium or thermochemical non-equilibrium gas models. Since the governing equations for the other models can typically be derived from the three-temperature thermochemical non-equilibrium governing equations, the latter are used as an example below to introduce the basic equations. For a gas mixture consisting of $ns$ species, the equations are as follows:

$$\frac{\partial Q}{\partial t} + \frac{\partial E}{\partial x} + \frac{\partial F}{\partial y} + \frac{\partial G}{\partial z} - \left(\frac{\partial E_v}{\partial x} + \frac{\partial F_v}{\partial y} + \frac{\partial G_v}{\partial z}\right) = S \qquad (1)$$

where $Q = \begin{bmatrix} \rho_1 \\ \vdots \\ \rho_{ns} \\ \rho u \\ \rho v \\ \rho w \\ \rho E \\ \rho e_v \\ \rho e_e \end{bmatrix}$, $E = \begin{bmatrix} \rho_1 u \\ \vdots \\ \rho_{ns} u \\ \rho u^2 + p \\ \rho uv \\ \rho uw \\ \rho u h_0 \\ \rho e_v u \\ \rho e_e u \end{bmatrix}$, $F = \begin{bmatrix} \rho_1 v \\ \vdots \\ \rho_{ns} v \\ \rho vu \\ \rho v^2 + p \\ \rho vw \\ \rho v h_0 \\ \rho e_v v \\ \rho e_e v \end{bmatrix}$, $G = \begin{bmatrix} \rho_1 w \\ \vdots \\ \rho_{ns} w \\ \rho wu \\ \rho wv \\ \rho w^2 + p \\ \rho w h_0 \\ \rho e_v w \\ \rho e_e w \end{bmatrix}$, $S = \begin{bmatrix} \omega_1 \\ \vdots \\ \omega_{ns} \\ 0 \\ 0 \\ 0 \\ 0 \\ \omega_v \\ \omega_e \end{bmatrix}$, $E_v =$

$\begin{bmatrix} J_{x1} \\ \vdots \\ J_{xns} \\ \tau_{xx} \\ \tau_{xy} \\ \tau_{xz} \\ u\tau_{xx} + v\tau_{xy} + w\tau_{xz} + q_x \\ +q_{vx} + q_{ex} + \rho \sum_{i=1}^{ns} D_i h_i \frac{\partial Y_i}{\partial x} \\ q_{vx} + \rho \sum_{i_{mol}} D_i e_v^i \frac{\partial Y_i}{\partial x} \\ q_{ex} + \rho \sum_{i=1}^{ns} D_i h_e^i \frac{\partial Y_i}{\partial x} \end{bmatrix}$, $F_v = \begin{bmatrix} J_{y1} \\ \vdots \\ J_{yns} \\ \tau_{yx} \\ \tau_{yy} \\ \tau_{yz} \\ u\tau_{yx} + v\tau_{yy} + w\tau_{yz} + q_y \\ +q_{vy} + q_{ey} + \rho \sum_{i=1}^{ns} D_i h_i \frac{\partial Y_i}{\partial y} \\ q_{vy} + \rho \sum_{i_{mol}} D_i e_v^i \frac{\partial Y_i}{\partial y} \\ q_{ey} + \rho \sum_{i=1}^{ns} D_i h_e^i \frac{\partial Y_i}{\partial y} \end{bmatrix}$, $G_v = \begin{bmatrix} J_{z1} \\ \vdots \\ J_{zns} \\ \tau_{zx} \\ \tau_{zy} \\ \tau_{zz} \\ u\tau_{zx} + v\tau_{zy} + w\tau_{zz} + q_z \\ +q_{vy} + q_{ey} + \rho \sum_{i=1}^{ns} D_i h_i \frac{\partial Y_i}{\partial z} \\ q_{vz} + \rho \sum_{i_{mol}} D_i e_v^i \frac{\partial Y_i}{\partial z} \\ q_{ez} + \rho \sum_{i=1}^{ns} D_i h_e^i \frac{\partial Y_i}{\partial z} \end{bmatrix}$.

In above, $\rho_i$ is the density of the $i$-th species, $Y_i$ is its mass fraction ($\rho_i/\rho$), and $i_{mol}$ represents index for polyatomic species. $\omega_i$, $\omega_v$ and $\omega_e$ denote the mass production rate, the vibrational energy source term, and the electronic energy source term, respectively; $E$, $e_v$ and $e_e$ represent the specific total energy, vibrational energy, and electronic energy, respectively. The mass diffusion of $i$-th species is $J_{x_j i} = \rho D_i \partial Y_i / \partial x_j$, with $D_i$ as the diffusion coefficient, and the specific total enthalpy is $h_0 = \sum_{i=1}^{ns} Y_i h_i + (\sum_{j=1}^{3} u_j^2)/2$, with $h_i$ as the specific enthalpy. The viscous stress is $\tau_{x_i x_j} = -(2/3)\mu \nabla \cdot \vec{V} + \mu(\partial u_i/\partial x_j + \partial u_j/\partial x_i)$, where $\mu$ is the viscosity coefficient. The pressure is $p = \sum_{i \neq i_e} \rho_i R_i T + p_e$, where $i_e$ indicates the electron, $R_i$ is the gas constant for the $i$-th species and $p_e$ is the electronic pressure. $(q, q_v, q_e)_{x,y,z}$ are the heat flux components in the three coordinate directions, calculated from $T_{tr}$, $T_v$ and $T_e$, respectively. Since the governing equations, thermodynamic relations, and transport coefficients differ among the various gas models, the details will be presented in Section 2.2.

2.2 Thermochemical equilibrium and non-equilibrium models

The gas models employed in this study include both equilibrium and non-equilibrium models. The latter encompass chemical non-equilibrium (Gupta 7 species and 9 reactions air model [17]) and thermal non-equilibrium models (i.e., *2T*- and *3T*-M). The *EGM* is introduced first in the following section.

2.2.1 Equilibrium gas model

For the *EGM*, this study employs the modified Tannehill curve-fit method developed by Srinivasan and Tannehill et al. in [18]. This method offers the advantages of relatively efficient implementation in computations. Since the *EGM* does not consider multiple species, the corresponding governing equations can be roughly derived from Eq. (1) by reducing $ns$ to 1, removing the vibrational and electronic energy conservation equations, replacing $\rho_i$ with $\rho$, and eliminating the five variables: $\omega_i$, $q_{x_j i}$, $\partial Y_i/\partial x_j$, $q_v$, and $q_e$. In the *EGM*, the fitted curves for thermodynamic variables or transport coefficients are composed of a series of piecewise functions defined over distinct intervals of input variables (e.g., $\rho$). A Grabau-type blending function is employed to connect these piecewise segments, whose general form is as below:

$$f(x,y) = f_1(x,y) + \frac{f_2(x,y) - f_1(x,y)}{1 \pm exp(k_0 + k_1 x + k_2 y + k_3 xy)} \tag{2}$$

where the "+" in the denominator denotes odd-function blending, the "−" denotes even-function blending, and the $k_i$ are constant coefficients. $f_1(x,y)$ and $f_2(x,y)$ are cubic polynomials defined as follows:

$$\begin{cases} f_1(x,y) = p_1 + p_2 x + p_3 y + p_4 xy + p_5 x^2 + p_6 y^2 + p_7 x^2 y + p_8 xy^2 + p_9 x^3 + p_{10} y^3 \\ f_2(x,y) - f_1(x,y) = p_{11} + p_{12} x + p_{13} y + p_{14} xy + p_{15} x^2 + p_{16} y^2 + p_{17} x^2 y + p_{18} xy^2 \\ \qquad\qquad\qquad + p_{19} x^3 + p_{20} y^3 \end{cases} \tag{3}$$

where $p_i$ here are constant coefficients.

By Eq. (2), the temperature can be fitted as $T = T(e, \rho)$ or $T = T(p, \rho)$. The effective specific heat ratio $\tilde{\gamma}$ can be fitted as $\tilde{\gamma} = \tilde{\gamma}(p, \rho)$. From these, the enthalpy is obtained as $h = (p/\rho)(\tilde{\gamma}/\tilde{\gamma} - 1)$ and the speed of sound as $a = \sqrt{\tilde{\gamma} R T}$. During the fitting process, the variables are first nondimensionalized and then log-transformed. Taking the $T = T(p, \rho)$ as an example, the fitting formula is $log_{10}(T/T_0) = f(Y, Z)$, where $X = log_{10}(p/p_0)$, $Y = log_{10}(\rho/\rho_0)$, $Z = X - Y$, with $T_0$ = 273.15 K, $\rho_0$ = 1.292 kg/m³, and $p_0$ = 1.0134×10⁵ N/m².

For the fitting of transport coefficients and $Pr$: $\mu = \mu(T, \rho)$ or $\mu = \mu(e, \rho)$, $k = k(e, \rho)$, and $Pr = Pr(T, \rho)$, with the note that the logarithmic transformation is only required when fitting is performed using $\rho$ and $e$. Taking the $\mu = \mu(T, \rho)$ as an example, the fitting formula is $\mu/\mu_0 = f(X, Y)$, where $X = T/1000$ K, $Y = log_{10}(\rho/\rho_0)$, $\mu_0$ = 16.5273×1.058×10⁻⁶ kg·s/m, and $\rho_0$ = 1.243 kg/m³. The detailed methodology can be found in [18].

2.2.2 Chemical non-equilibrium

Chemical reactions of finite rate are considered in chemical non-equilibrium gas model (CNEG). If there are $ns$ species and $nr$ reversible reactions, the reactions can be as follows:

$$\sum_{i=1}^{ns} \alpha_{ji} A_i \rightleftharpoons \sum_{i=1}^{ns} \beta_{ji} A_i \quad (j = 1, \cdots, nr) \tag{4}$$

where $\alpha_{ji}$ and $\beta_{ji}$ are stoichiometric coefficients, those for colliders are zero, and $A_i$ denotes the species $i$. The $\omega_i$ in Eq. (1) is defined as: $\omega_i = M_i \sum_{j=1}^{nr} (d(\rho_i/M_i)/dt)_j$, where $M_i$ is the molecular weight, and the change rate of molar concentration in the $j$-th reaction is given by:

$$\left(d(\tfrac{\rho_i}{M_i})/dt\right)_j = (\beta_{ji} - \alpha_{ji}) \left[ k_{f,j} (\prod_{m=1}^{ns} \left(\tfrac{\rho_m}{M_m}\right)^{\alpha_{jm}}) \left(\sum_{k=1}^{ns} \tfrac{\rho_k}{M_k} C_{kj}\right)^{L_j} - k_{b,j} (\prod_{m=1}^{ns} \left(\tfrac{\rho_m}{M_m}\right)^{\beta_{jm}}) \left(\sum_{k=1}^{ns} \tfrac{\rho_k}{M_k} C_{kj}\right)^{L_j} \right] \tag{5}$$

where $C_{kj}$ is the three-body collision coefficient of species $k$ (note that the same coefficient applies to both the forward and backward reaction directions), $L_j$ is a switch indicating whether to consider the three-body collision, and the forward and backward reaction rate coefficients, $k_{f,j}$ and $k_{b,j}$, are given by the Arrhenius formula:

$$k_{f/b,j} = A_{f/b,j} T_c^{B_{f/b,j}} e^{-E_{f/b,j}/(R_u T_c)} \tag{6}$$

where the subscript "$f/b$" denotes the forward/backward reaction, $A_{f/b,j}$, $B_{f/b,j}$, $E_{f/j,j}$, $R_u$ are reaction-specific coefficients, and $T_c$ is the controlling temperature. This study adopts the air model with 7 species and 9 reactions proposed by Gupta [17], as listed in Table 1. $M_{1\sim 4}$ denote colliders. The detailed coefficients for reactions can be found in [17].

Table 1. The air chemistry model with 7 species and 9 reactions [17]

| Index | Reaction equation | Index | Reaction equation |
|---|---|---|---|
| 1 | $O_2 + M_1 \Longleftrightarrow 2O + M_1$ | 6 | $N_2 + O \Longleftrightarrow NO + N$ |
| 2 | $N_2 + M_2 \Longleftrightarrow 2N + M_2$ | 7 | $N + O \Longleftrightarrow NO^+ + e^-$ |
| 3 | $N_2 + N \Longleftrightarrow 2N + N$ | 8 | $O_2 + N_2 \Longleftrightarrow NO + e^- + NO^+$ |
| 4 | $NO + M_3 \Longleftrightarrow N + O + M_3$ | 9 | $NO + M_4 \Longleftrightarrow NO^+ + e^- + M_4$ |
| 5 | $NO + O \Longleftrightarrow O_2 + N$ | | |

| $M_1 = N, O, O_2, N_2, NO; M_2 = O, O_2, N_2, NO; M_3 = O, N, O_2, N_2, NO; M_4 = O_2, N_2$ |
|---|

Additionally, the air model with 7 species and 11 reactions proposed by Park et al. [19] also employed in this study for some cases, which is illustrated in Table 2 with details shown in [19].

Table 2. The air chemistry model with 7 species and 11 reactions [19]

| Index | Reaction equation | Index | Reaction equation |
|---|---|---|---|
| 1 | $O_2 + M_1 <=> 2O + M_1$ | 7 | $N_2 + e^- <=> 2N + e^-$ |
| 2 | $O_2 + M_2 <=> 2O + M_2$ | 8 | $NO + M_4 <=> N + O + M_4$ |
| 3 | $N_2 + M_3 <=> 2N + M_3$ | 9 | $NO + O <=> N + O_2$ |
| 4 | $N_2 + N <=> 2N + N$ | 10 | $N_2 + O <=> NO + N$ |
| 5 | $N_2 + O <=> 2N + O$ | 11 | $N + O <=> NO^+ + e^-$ |
| 6 | $N_2 + NO <=> 2N + NO$ | | |
| $M_1 = N, O; M_2 = O_2, N_2, NO, e^-; M_3 = O_2, N_2; M_4 = O_2, N_2, N, O, NO, e^-$ | | | |

When the above reaction mechanism is combined with the *1T*-M, the one-temperature CNEG model is formed, which is still referred to as *1T*-M next for brevity. Its governing equations can be obtained by removing the vibrational and electronic energy equations as well as the variables $q_v$ and $q_e$ from Eq. (1). Furthermore, in Eq. (6), $T_c$ is set equal to $T$.

2.2.3 Multi-temperature models

For multi-temperature models, the excitation of vibrational and electronic energies in gas molecules is considered. These internal energy modes are typically not in equilibrium with the translational-rotational one, necessitating the use of separate $T_v$ or $T_e$ to describe the energy state. Depending on the extent of energy mode excitation, common models include the *2T*-M (featuring $T_{tr}$ and $T_{ve}$), the *3T*-M (featuring $T_{tr}$, $T_v$ and $T_e$), as well as models considering more temperatures. Since the *2T*-M can be formally derived from the *3T*-M by setting $T_v = T_e = T_{ve}$, the *3T*-M and its associated formulation are introduced first in the following.

(1) Three-temperature model [20]

For the *3T*-M, the complete governing equations are given by Eq. (1), where the vibrational energy source term $\omega_v$ is expressed as:

$$\omega_v = \sum_{i_{mol}} \rho_i \frac{e_{v,i}^* - e_{v,i}}{\tau_{v,i}} + \sum_{i_{mol}} \frac{\rho_i}{M_i} \frac{e_{v,i}^{**} - e_{v,i}}{\tau_{e,i}} + \sum_{i_{mol}} \omega_i D_i \tag{7}$$

where the three terms on the right-hand side represent the energy exchange between the translational and vibrational modes, the energy exchange between the electronic and vibrational modes, and the vibrational energy gain/loss due to molecular dissociation and atomic recombination, respectively. In Eq. (7), $e_{v,i}^*$ denotes the molar vibrational energy of the $i$-th species evaluated at $T_{tr}$, while $e_{v,i}^{**}$ denotes the molar vibrational energy evaluated at $T_e$. The term $\tau_{v,i}$ represents the translational-vibrational relaxation time of the $i$-th species, and $\tau_{e,i}$ represents the corresponding electron-vibrational relaxation time. Their specific calculation formulas can be found in [21, 22].

The electronic energy source term $\omega_e$ is given by [23]:

$$\omega_e = 2\rho_e \frac{3}{2} R(T_{tr} - T_e) \sum_{i \neq i_e} \frac{v_{e,i}}{M_i} + - \sum_{i_{mol}} \frac{\rho_i}{M_i} \frac{e_{v,i}^{**} - e_{v,i}}{\tau_{e,i}} - (R_O I_O + R_N I_N) \tag{8}$$

where the three terms on the right-hand side represent, respectively, the energy exchange between heavy particles and the electron in translational mode, the energy exchange between vibrational and electronic modes, and the energy loss due to electron-impact dissociation. In Eq. (8), $v_{e,i}$ represents the effective collision frequency between electrons and heavy particles of the $i$-th species. $R_O$ and $R_N$ denote the forward reaction rates of the forced ionization reactions for O and N, respectively, while $I_O$ and $I_N$ are the molar first ionization energies of O and N, respectively. Detailed descriptions of the parameters involved in the above formulas can be found in [23].

(2) Two-temperature model [24]

In the *2T*-M, it is assumed that $T_v = T_e$, and they are collectively denoted as $T_{ve}$. The corresponding vibrational-electronic energy equation can be derived by summing the vibrational and electronic energy equations in Eq. (1) as:

$$\frac{\partial(\rho e_{ve})}{\partial t} + \frac{\partial(\rho e_{ve} u_j)}{\partial x_j} = \partial\left(k_{ve}\frac{\partial T_{ve}}{\partial x_j}\right)/\partial x_j + \partial\left(\sum_{i=1}^{ns}\rho D_i\left(\frac{e_{ve,i}}{M_i}\right)\frac{\partial Y_i}{\partial x_j}\right)/\partial x_j + \omega_{ve} \qquad (9)$$

where the specific vibrational-electronic energy is $e_{ve} = e_v + e_e$, the vibrational-electronic energy source term is $\omega_{ve} = \omega_v + \omega_e$, and $k_{ve}$ is the thermal conductivity for the *2T*-M.

The detailed parameter determination for the *2T*- and *3T*-M can be found in [20, 24]. Hypersonic thermal non-equilibrium is typically coupled with chemical non-equilibrium and must be accounted for in the computations. One implementation approach is through the $T_c$ shown in Eq. (6) [24]. Specifically, for the *2T*-M, $T_c$ can be expressed as:

$$T_c = T_{tr}^{a_c} T_{ve}^{b_c} \qquad (10)$$

where the exponents $a_c$ and $b_c$ take different values for different reaction types. For instance, in collisional dissociation or exchange reactions, $a_c = b_c = 0.5$. For the *3T*-M, following the suggestion of Kim et al. [22], $T_c$ for chemical reactions is defined as:

$$T_c = T_{tr}^{a_c} T_v^{b_c} T_e^{c_c} \qquad (11)$$

where the specific values of the exponents $a_c$, $b_c$, and $c_c$ can be found in [22].

In this study, the uses of multi-temperature models are integrated with CNEG as in *1T*-M, which are also referred to as *2T/3T*-M for brevity.

2.3 Boundary conditions

Two types of wall boundary conditions are employed herein: non-slip and slip boundaries. The wall temperature of the cone is set to $T_w$ = 1250 K. A brief introduction is provided below.

(1) Non-slip boundary condition

The corresponding conditions include: zero velocity at the wall, zero normal pressure gradient, and an isothermal wall. For the multi-temperature models, the temperatures are set as $T_{tr,w} = T_{ve,w} = T_{v,w} = T_{e,w} = T_w$. Specifically, for the *EGM*, $\rho_w$ needs to be solved iteratively using Newton's method: $\rho_w^{k+1} = \rho_w^k + (T_w - T(p_w, \rho_w^k))/(\partial T/\partial \rho)_p^k$, where $k$ is the iteration step and the relation between $T$ and $\rho$ adopts the aforementioned fitting formula $T(p,\rho)$.

(2) Slip boundary condition

Since the altitude of freestream considered herein is 72 km, it is necessary to compare and analyze the potential influence of slip flow. To balance between computational cost and model fidelity, the Maxwell–Smoluchowski (M–S) slip boundary condition is additionally employed for the *3T*-M cases. The Maxwell velocity slip condition [25] is given by:

$$u_s = u_w + ((2 - \sigma_u)/\sigma_u)\lambda\,(\partial u_\tau/\partial n) \qquad (12)$$

where $u_s$ is the slip velocity at the wall, $u_\tau$ is the tangential velocity, $u_w$ is the wall velocity (zero in current study), $n$ is the wall-normal direction, $\sigma_u$ is the momentum accommodation coefficient (taken as 1 herein), and $\lambda$ is the mean free path. The jump of $T_{tr}$ is described by the Smoluchowski temperature jump condition [26]:

$$T_{tr,s} - ((2 - \sigma_{T_{tr}})/\sigma_{T_{tr}})(2\gamma/Pr(\gamma + 1))\lambda(\partial T_{tr}/\partial n) = T_w \qquad (13)$$

where $T_{tr,s}$ is the slip $T_{tr}$ at the wall, and $\sigma_{T_{tr}}$ is corresponding accommodation coefficient (taken as 1 here). The slip $T_v/T_e$ is addressed through the vibrational/electronic energy jump condition:

$$e_{v/e,s} - e_{v/e,w} = ((2 - \sigma_{v/e})/\sigma_{v/e})\lambda(\partial e_{v/e}/\partial n) \qquad (14)$$

where $e_{v/e,s}$ is the slip vibrational/electronic energy at the wall, $e_{v/e,w}$ is the energy calculated by $T_w$, and $\sigma_{v/e}$ is the corresponding accommodation coefficient (taken as 0.001 for both modes). Due to the small value of $\sigma_{v/e}$, the boundary condition of $T_v/T_e$ approximates that of an adiabatic wall with regard to the corresponding temperature.

2.4 WENO3-$Z_{ES3}$ scheme

To enhance the accuracy of the simulations, WENO3-$Z_{ES3}$ [16] is adopted in this study. As demonstrated in [16], this scheme offers improved resolution and lower numerical dissipation while maintaining well robustness. The following uses the one-dimensional hyperbolic equation $u_t + f(u)_x = 0$ where $\partial f(u)/\partial u > 0$ for illustration. The conservative discretization of $f(u)_x$ at $x_j$ can be expressed as $\partial f(u)/\partial x_j \approx (\hat{f}_{j+1/2} - \hat{f}_{j-1/2})/\Delta x$, where $\hat{f}_{j\pm 1/2}$ denotes the numerical flux. Assuming $r$ represents the number of sub-stencils as well as the number of grid points per stencil, WENO scheme with an order of $r$ can be formulated as $\hat{f}_{j+1/2} = \sum_{k=0}^{r-1} \omega_k q_k^r$ where the linear interpolation on the sub-stencil is $q_k^r = \sum_{l=0}^{r-1} a_{kl}^r f(u_{j-r+k+l+1})$ with $a_{kl}^r$ as the interpolation coefficients. The nonlinear weights $\omega_k$ are obtained by the unnormalized weight $\alpha_k$ as $\omega_k = \alpha_k / \sum_{l=0}^{r-1} \alpha_l$. For the WENO-Z type scheme, $\alpha_k$ may take the form $\alpha_k = d_k \left(1 + c_\alpha \left(\tau/(\beta_k^{(r)} + \varepsilon)\right)^p\right)$ where $d_k$ are linear weights, $\beta_k^{(r)}$ and $\tau$ are the local and global smoothness indicator respectively, $p = 1$ or $2$, $c_\alpha$ may take 1, and $\varepsilon$ can take a small value such as $10^{-40}$. A detailed accuracy analysis was provided in [16] on the third-order schemes, and a novel critical point accuracy analysis was proposed to address the order reduction problem near critical points, namely a critical point being located at any position within the interval rather than only at $x_j$. Based on this understanding, [16] proposed a new $\tau_4$ and $\beta_k$ variants. The final expression for $\alpha_k$ is as follows: $\alpha_k = d_k \left(1 + c_\alpha \left(\tau_4/(\beta_k^{(r)*} + \varepsilon)\right)^2\right)$, where $c_\alpha = 0.4$, and $\tau_4 = |(2f_{j+1} - 3f_j + f_{j-1})(f_{j+2} - 3f_{j+1} + 3f_j - f_{j-1})|$, $\beta_0^{(2)*} = (f_j - f_{j-1})^2 + c_{\beta_0}(f_j - 2f_{j-1} + f_{j-2})^2$, $\beta_1^{(2)*} = (f_{j+1} - f_j)^2 + c_{\beta_1}(f_{j+2} - 2f_{j+1} + f_j)^2$. The coefficients are $c_{\beta 0} = 0.5$ and $c_{\beta 1} = 0.15$. Further details of the scheme are suggested to [16].

2.5 Validation tests

Since comprehensive validations for the *EGM*, the M-S boundary condition, and WENO3-$Z_{ES3}$ have been presented separately in [27], [28], and [16], respectively, they are not repeated here for brevity. This section primarily focuses on the verification of the thermochemical non-equilibrium models described in Section 2.2. For this purpose, two cases—specifically, the hypersonic flow over the ELECTRE spherically blunted cone and the RAM-C blunt cone—are selected, and simulations are conducted under non-slip boundary conditions.

(1) Hypersonic flow over the ELECTRE spherically blunted cone at $M_\infty = 12.9$

The freestream conditions are $M_\infty = 12.9$, $T_\infty = 265\,\text{K}$, $T_w = 343\,\text{K}$, and $Re/m = 1.63\times 10^5$ /m. The simulations employ the *1T*-, *2T*- and *3T*-M with the reaction model chosen as the one in Table 1. The freestream mass fractions are $Y_{N_2,\infty} = 0.767$ and $Y_{O_2,\infty} = 0.233$.

Fig. 1 presents a comparison of distributions of the wall pressure coefficient ($C_P$) and wall heat flux ($Q_w$) by different temperature models, as well as that of temperatures and $Y_i$ along the stagnation line. The reference data for temperatures and species are calculated by HYFLOW [29]

(which has only *1T*- and *2T*-M prediction capabilities). In the figure as well as the subsequent figures and subscripts of symbol, the notations "*1T*", "*2T*", and "*3T*" denote predictions using the *1T*-, *2T*-, and *3T*-M under non-slip boundary conditions, respectively; meanwhile, the notation "*EG*" concerns the *EGM* with non-slip conditions while "*3T$_{slip}$*" regards the *3T*-M with slip conditions.

The results in Fig. 1(a) and (b) show that the $C_P$ and $Q_w$ distributions from the different temperature models agree well with the flight data [30], particularly the $C_P$ distribution from the *3T*-M. Fig. 1(c) shows that the results of the *1T*- and *2T*-M are generally consistent with the reference results, and the distributions of $T_{tr,2T}$ and $T_{ve,2T}$ are largely consistent with those of $T_{tr,3T}$ and $T_{v,3T}$, respectively. This consistency indicates the predictive capability of the present multi-temperature models. Notably, the distribution of $T_{e,3T}$ differs from that of $T_{v,3T}$, indicating the significant excitation of non-equilibrium electronic state in this case. Fig. 1(d) shows that $Y_i$ from the *1T*- and *2T*-M agree well with the reference data. The distribution trends of $Y_i$ from different temperature models are generally consistent. While the differences between the *2T*- and *3T*-M results are relatively minor, both exhibit certain discrepancies in $Y_i$ distributions compared to the *1T*-M due to the different evaluation of $T_c$.

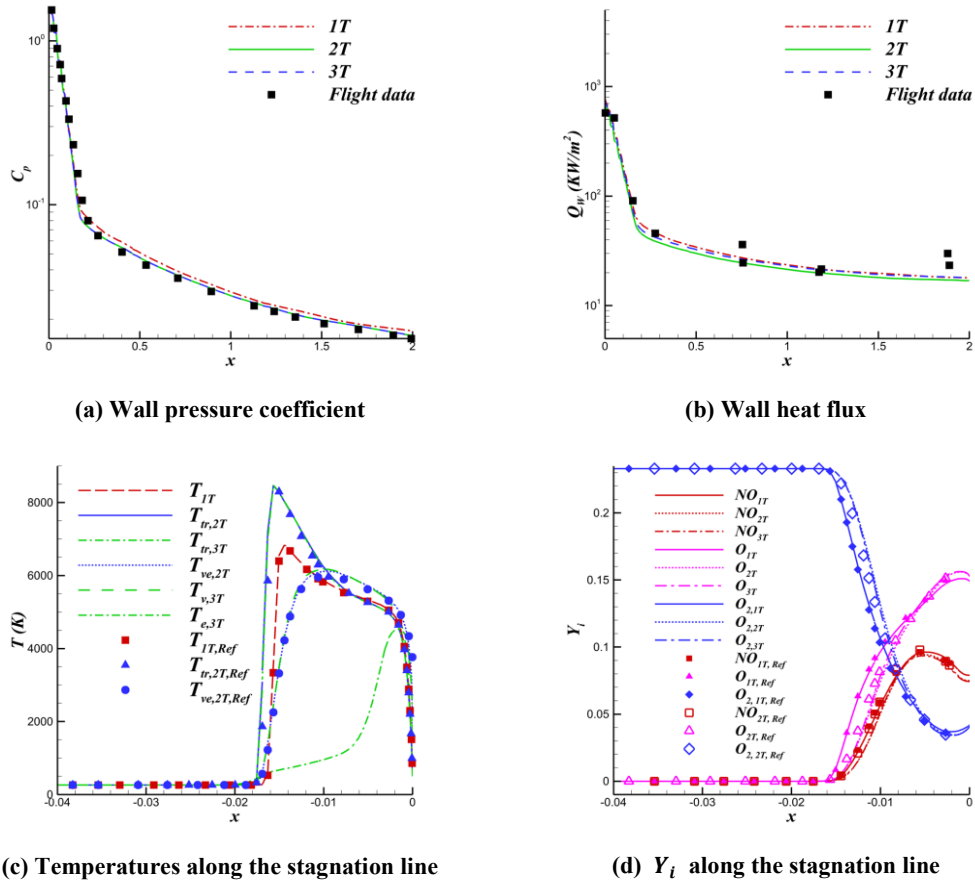

(a) Wall pressure coefficient  (b) Wall heat flux

(c) Temperatures along the stagnation line  (d) $Y_i$ along the stagnation line

**Fig. 1 Wall and stagnation-line distributions of flow properties compared with flight data [30] and predictions by HYFLOW [29]**

(2) Hypersonic flow over the RAM-C blunted cone at $M_\infty$ = 24.0

To further validate the implementation of the *3T*-M, simulations are performed for the RAM-C blunt cone. The freestream conditions are: $H$ = 61 km, $M_\infty$ = 24.0, $\rho_\infty$ = 2.73×10$^{-4}$ kg/m³, $T_\infty$ = 244 K, and $T_w$ = 1500 K. The reaction model is specified in Table 2, with freestream $Y_i$ as $Y_{N_2,\infty}$ = 0.23 and $Y_{O_2,\infty}$ = 0.77. 错误!未找到引用源。(a) compares the temperature distributions along the stagnation line of the blunt cone against reference results [31] using *2T*-M. It indicates that $T_{tr,3T}$ and $T_{v,3T}$ agree well with the reference values. While $T_{e,3T}$ follows a similar distribution pattern to $T_{v,3T}$, their magnitudes differ. 错误!未找到引用源。 (b) shows the peak of electron number density along the wall-normal direction. The results are consistent in both trend and order of magnitude with the reflectometer measurements in [32].

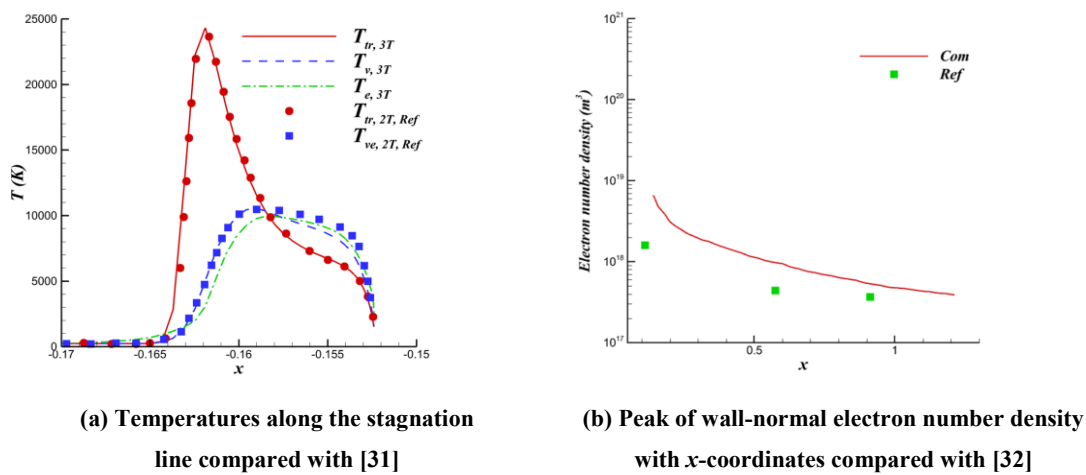

(a) Temperatures along the stagnation line compared with [31]

(b) Peak of wall-normal electron number density with *x*-coordinates compared with [32]

**Fig. 2 Distributions of various temperatures on stagnation-line and peak of wall-normal electron number density with *x*-coordinates**

In short, the validation cases mentioned indicate the validation of the physicochemical models and numerical scheme employed in this study.

## 3 Comparative study of equilibrium and non-equilibrium characteristics predicted by different models

Vehicles flying at high-altitude and high-*Ma* encounter an extreme thermochemical non-equilibrium environment. Predictions of flow fields and aerodynamic characteristics may differ among various equilibrium and non-equilibrium models, rendering a comparative analysis of significant engineering value. This section focuses on a conical configuration, and continuum flow simulations under non-slip boundary conditions are first performed using the *EGM*, *1T*-, *2T*-, and *3T*-M, respectively. Given the consideration of altitude as 72 km in computation, slip-flow calculations incorporating the M-S slip boundary condition are also conducted based on the *3T*-M for comparison. The study places particular emphasis on the equilibrium and non-equilibrium flow characteristics predicted by the different models, especially the properties of the base flow. The computational conditions and grid convergence study are introduced below.

3.1 Numerical setup and grid convergence study

The freestream conditions are as follows: $M_\infty$ = 27, $T_\infty$ = 211.926 K, Re/m = 37580.1 /m, and $T_w$ = 1250 K. The angles of attack (AOAs) for the computations are $\alpha$ = 0° and 8°, where $\alpha$ is defined as the angle between the cone axis and the freestream velocity vector in the symmetrical meridional plane (*SMP*), while the sideslip angle is zero. For thermochemical equilibrium，the *EGM* is adopted; while for non-equilibrium simulations, the *1T*-, *2T*-, and *3T*-M described in Section 2 are employed with the integration with the Gupta 7 species and 9 reactions air reaction model. The cone geometry is illustrated in Fig. 3, featuring a nose radius of 10 mm, a half-cone angle $\theta$ = 10°, and an axial length $L$ = 4500 mm.

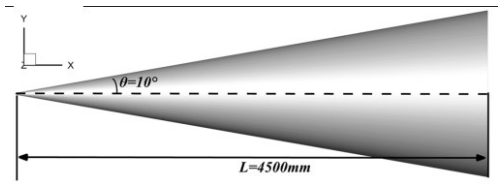
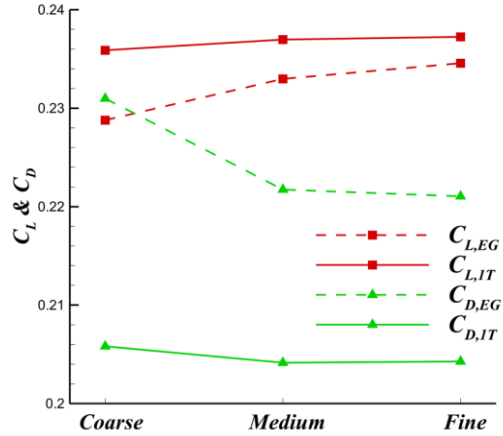

Fig. 3 Schematic of the cone geometry

Fig. 4 Lift and drag coefficients computed with different grids

To determine an appropriate grid, the grid convergence study was conducted. For the non-slip continuum flow, three structured grids of varying resolution were generated: Coarse (1.0 million cells), Medium (2.0 million cells), and Fine (2.9 million cells). The normal spacing of the first grid layer for all grids is $6\times10^{-6}$ m. The *EGM* and *1T*-M were selected as representatives to perform the grid convergence study at $\alpha=8°$. The lift coefficient ($C_L$) and drag coefficient ($C_D$) obtained from grids of different resolutions are presented in Fig. 4. The results show that the values computed from the Medium and Fine grids are in close agreement, while those from the Coarse grid exhibit discernible discrepancies. This indicates that the Medium grid provides a grid-converged solution. Therefore, it was selected for all subsequent computations. Building upon this, the normal spacing of the first grid layer at the wall was appropriately increased for the slip-flow cases, based on the $\lambda$ near the wall computed using the *1T*-M.

3.2 Analysis of overall flow structure and aerodynamic characteristics

This subsection investigates the flow structure around the hypersonic cone using different models by analyzing the pressure and *Ma* distributions at two AOAs. A comparative analysis of the aerodynamic coefficients and wall parameter distributions is also conducted. Since the flow structure over the cone forebody is relatively straightforward, the analysis places more emphasis on the base flow. To conserve space in the figures, results on the horizontal plane ($y = 0$) for $\alpha=0°$ are often divided into two regions using the $z = 0$ line as a boundary to display different results (e.g., Fig. 5). For $\alpha=8°$, results are sometimes split into left and right regions using the $z = 0$ section as the boundary (e.g., Fig. 8).

3.2.1 Flow structure

（1）Case of $\alpha=0°$

At $\alpha = 0°$, the flow field around the cone exhibits an overall axisymmetric structure, and the pressure and *Ma* distributions are similar across the different models. Therefore, Fig. 5 uses the results from the *2T*-M as an example, presenting the pressure and *Ma* contours along with their iso-lines on the cone's horizontal plane ($y = 0$), where the pressure is normalized as $P^* = P/\rho U_\infty^2$. The inset in the figure highlights the overall flow structure, such as the conical shock close to the cone surface. It can be observed that a conical shock exists along the cone body, with a radial expansion wave (*EW*) structure extending downstream in the direction of the conical surface, followed by a recompression wave (*RCW*). Furthermore, a pressure maximum is present at the center of the base region. Fig. 6 presents the *Ma* and streamline distributions on the horizontal plane ($y = 0$) for the

*EGM* and *2T*-M, respectively, where the solid white line denotes the sonic line (*SNL*). The results show that flow separation occurs at the outer edge of the cone base, leading to the formation of a vortex within the base region. Furthermore, the *SNL* qualitatively coincides with the separating streamline during the initial and intermediate stages of separation. A comparison between models reveals that the flow field predicted by the *EGM* exhibits a greater streamwise extent, while that from the *2T*-M possesses a larger overall radial scale.

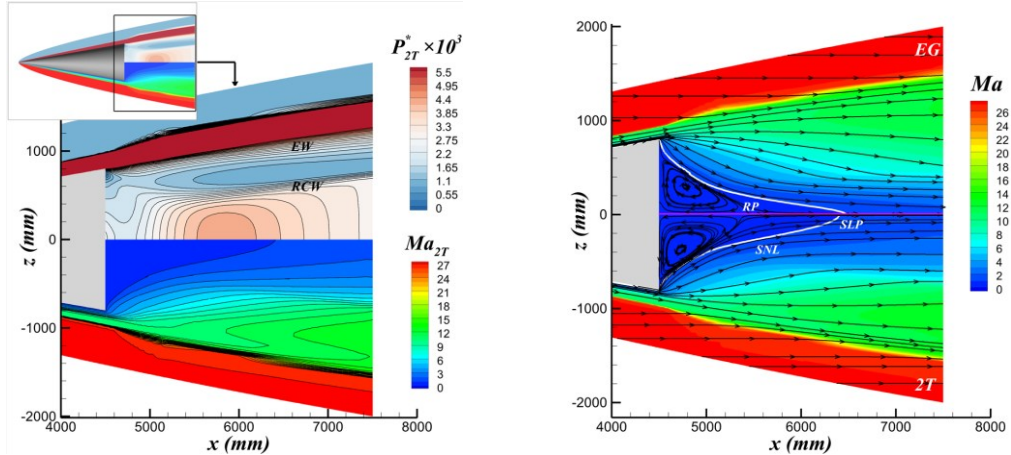

Fig. 5 Pressure (up) and *Ma* (down) contours and iso-lines on the cone horizontal plane ($y = 0$) by *2T*-M at $\alpha = 0°$

Fig. 6 *Ma* contours and streamline distributions on the cone base horizontal plane ($y = 0$) by *EGM* (up) and *2T*-M (down) at $\alpha = 0°$

To quantitatively illustrate the differences predicted by the various models in characteristics such as the extent of the subsonic region and the location of the flow reattachment point (*RP*), Fig. 7 comprehensively presents the pressure extrema ($P^*_{EXT}$), and *x*-coordinates of *RP*, intersection points of the *SNL* with the centerline (*SLP*), and locations of $P^*_{EXT}$. As mentioned above, notations such as *EGM* and 1*T*-M are simplified as *EG* and 1*T* respectively in the figure as well as in the subscript of symbol. The results show that compared to results of non-equilibrium gas models, the *EGM* yields little variation in the measured locations except for a larger $x_{P^*_{EXT}}$, and its $P^*_{EXT}$ is relatively smaller. Among results of three non-equilibrium models, the $x_{RP}$ values are similar, the same applies to $x_{SLP}$ and $x_{P^*_{EXT}}$. The $P^*_{EXT}$ follow the order $P^*_{EXT,1T} > P^*_{EXT,2T} > P^*_{EXT,3T}$. For the *3T$_{slip}$*-M, the values of $x_{RP}$, $x_{SLP}$, and $x_{P^*_{EXT}}$ are close to those of the *3T*-M, while its $P^*_{EXT}$ is comparatively smaller.

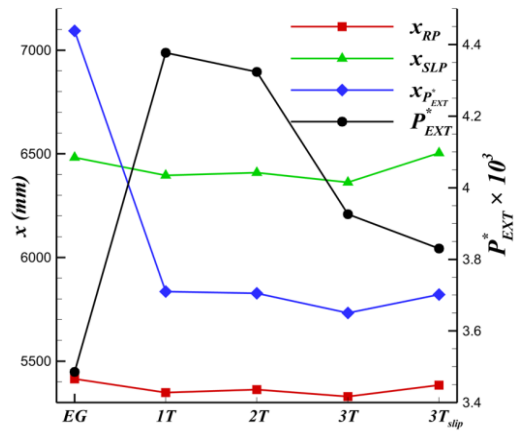

**Fig. 7 Streamwise locations of the *RP*, *SLP*, and pressure extremum point, along with values of the $P^*_{EXT}$ predicted by different models**

(2) Case of $\alpha=8°$

Fig. 8 first presents the cross-section pressure contour and iso-surface (corresponding to $P^*_{2T}$ = 4×10⁻³ and $P^*_{3T}$ = 3.7×10⁻³) distributions in the base region at $\alpha$ = 8°, obtained from different models. The results indicate that the overall pressure structures are similar across the models, with two distinct pressure maxima structures present in all cases (labeled "1" and "2" in the figure). For quantitative comparison, Fig. 9 presents the two $P^*_{EXT}$ and their corresponding locations under the different models. To save space, the figure also includes the coordinates of RP, as indicated by the sectional streamlines on the SMP (see Fig. 11). The subscripts "1" and "2" denote the extremum points labeled in Fig. 8. The comparisons show that: (a) Compared to the non-equilibrium gas, the location of $P^*_{EXT,EG}$ is farther from both the base surface and the centerline, and their values follow $P^*_{EXT,EG} < P^*_{EXT,1T}$; (b) For the three non-equilibrium models, the values of $x_{P^*_{EXT1,2}}$ show little variation, and a similar trend is observed for $y_{P^*_{EXT2}}$ and $x_{RP}$. The magnitude relationships of $y_{RP}$ across models resemble those of $y_{P^*_{EXT1}}$, both following $y_{1T} > y_{2T} > y_{3T}$. The quantity of $P^*_{EXT}$ satisfy $P^*_{EXT,1T} > P^*_{EXT,2T} > P^*_{EXT,3T}$; (c) For the $3T_{slip}$-M, $y_{RP}$ and $y_{P^*_{EXT1}}$ are slightly larger than those of the 3T-M, while the other locations show minimal differences. Additionally, the $P^*_{EXT}$ values of the former are slightly larger.

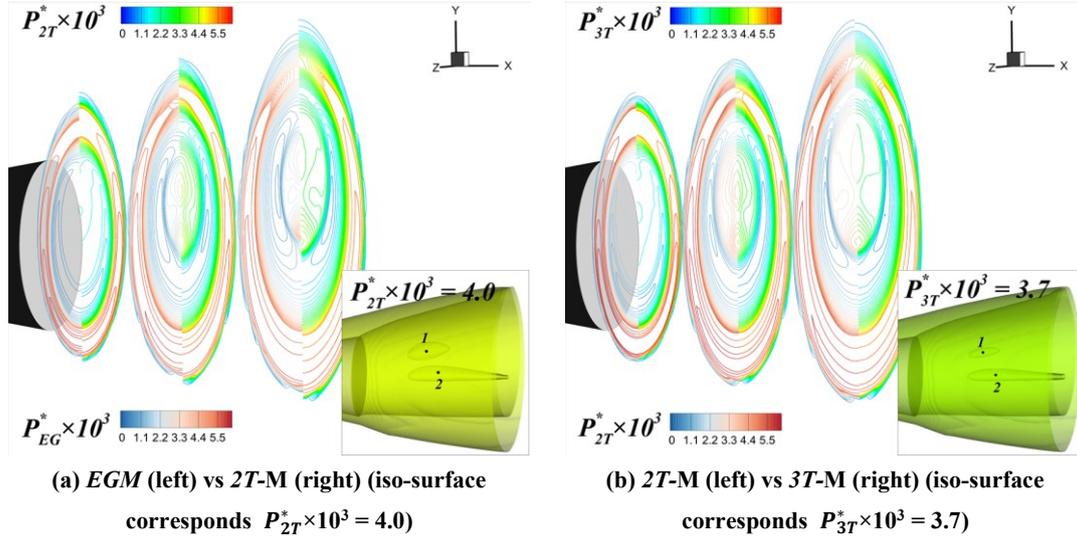

(a) EGM (left) vs 2T-M (right) (iso-surface corresponds $P^*_{2T}$×10³ = 4.0)

(b) 2T-M (left) vs 3T-M (right) (iso-surface corresponds $P^*_{3T}$×10³ = 3.7)

ig. 8 Pressure contours on streamwise cross-sections and iso-surfaces at cone base by different models at $\alpha$ = 8°

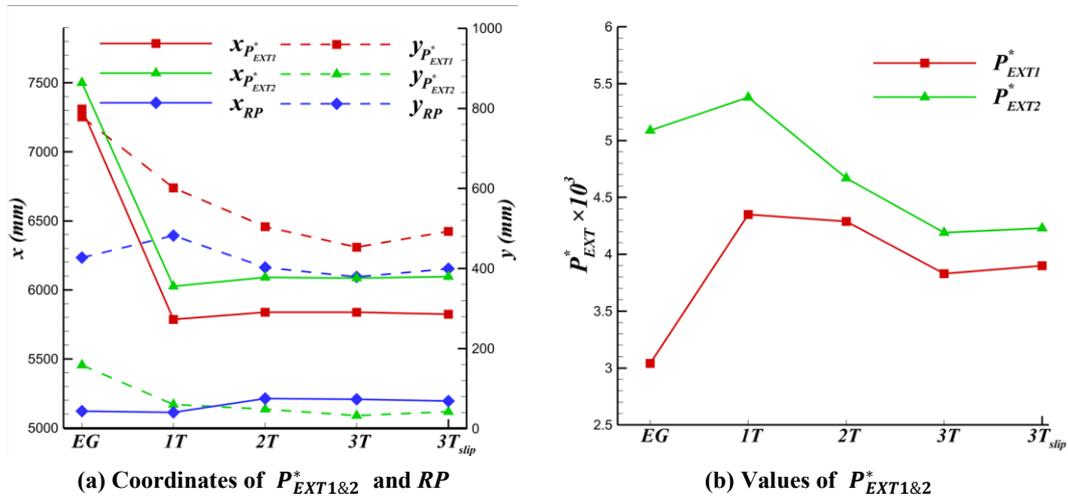

(a) Coordinates of $P^*_{EXT1\&2}$ and RP

(b) Values of $P^*_{EXT1\&2}$

Fig. 9 Coordinates of $P^*_{EXT1\&2}$ and RP at the SMP and values of $P^*_{EXT1\&2}$ at the cone base by different

Fig. 10 presents the *Ma* contours on the *SMP* as well as streamwise cross-sectional iso-lines, along with $Q_w$ as predicted by the *2T*-M at $\alpha = 8°$. An inset is provided, showing the cone wall pressure ($P_w$), streamlines, and the streamwise cross-sectional pressure iso-lines to illustrate the overall flow structure. It can be observed that an extremum of $Q_w$ exists at the base, with its maximum shifted leeward in the presence of $\alpha$. To reveal the corresponding flow structure, Fig. 11 presents the streamlines on base wall and those on several cross-sections (including the *SMP*) from the *2T*-M. These sections are perpendicular to the base with intersections indicated by dashed lines (see the upper-right inset). The results show that the flow topology on base wall consists of a primary node ($N_1$) and two primary saddle points ($S_1$ and $S_3$). Along the separating streamline (*SL*), secondary saddle points ($S_2$) and nodes ($N_2$ and $N_3$) appear alternately. The measured angles are approximately: $\angle N_2 N_1 S_1 \approx 9°$, $\angle S_2 N_1 S_1 \approx 29°$, and $\angle N_3 N_1 S_3 \approx 6°$. The surface separation topology is identified as follows: the primary reattachment of separation on the base wall occurs at $N_1$. The separation originates from $S_1$, $S_3$, and $S_2$, and terminates at $N_2$ and $N_3$. It is noteworthy that the topology includes a separation origin at $S_2$, a feature rarely observed or explicitly identified in the typical literature. Close analysis further reveals that the *SL* exhibits a quasi-circular structure, centered at coordinates ($y, z$) = (110, 0) mm with a radius of approximately 650 mm. The cross-sectional streamlines indicate the presence of an asymmetric vortex ring at the base, characterized by a smaller upper part and a larger lower part.

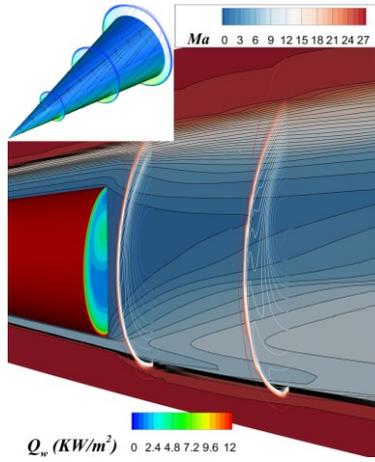
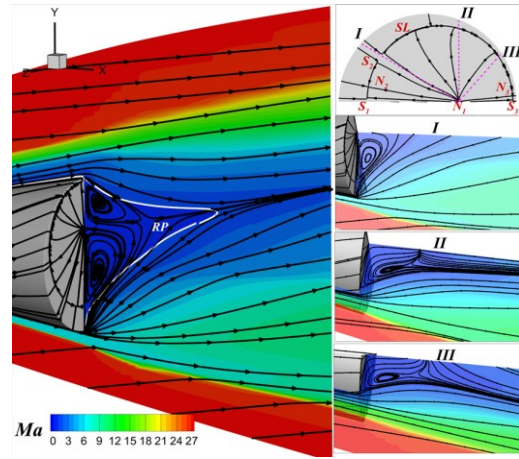

Fig. 10 *Ma* contours on the *SMP* with streamwise cross-sectional iso-lines and the base wall heat flux by *2T*-M at $\alpha = 8°$, where the inset illustrates wall pressure and streamlines with the pressure streamwise cross-sectional iso-lines

Fig. 11 Distributions of surface streamlines and selected sectional streamlines overlaid on a background of *Ma* contours by *2T*-M at $\alpha = 8°$, where the upper-right inset illustrates the locations of sub-sections I–III with purple dashed lines

3.2.2 Analysis of aerodynamic characteristics

To comparatively analyze the aerodynamic characteristics predicted by different models, the $C_D$ at both AOAs and the $C_L$ at $\alpha = 8°$ are presented. Furthermore, a decomposition of the $C_D$ into the pressure drag coefficient ($C_{D,p}$) and the skin friction drag coefficient ($C_{D,f}$) is provided, as shown in Fig. 12. The results are as follows: (a) At $\alpha = 8°$, the $C_L$ predicted by the various models are all very close; (b) The equilibrium gas yields larger $C_D$ and $C_{D,f}$ compared to the non-equilibrium gases, while their $C_{D,p}$ values are similar. For the different temperature models, the values of $C_D$ satisfy $C_{D,1T} \approx C_{D,2T} > C_{D,3T}$. The $C_{D,f}$ follows the same trend as $C_D$, whereas the $C_{D,p}$ again shows little variation. This indicates that the differences in $C_D$ among the models primarily stem from $C_{D,f}$; (c) At both AOAs, all coefficients show negligible differences between the *3T*- and *3T$_{slip}$*-M. Furthermore, taking $C_{D,3T}$ as the reference, the relative errors of the *EGM*, *1T*-, and *2T*-M are 11.51%, 5.03%, and 5.08% at $\alpha = 0°$, and 14.38%, 5.31%, and 5.82% at $\alpha = 8°$. Fig. 13 presents the center-of-pressure location ($X_{cp}$) and pitching moment (*PM*) at $\alpha = 8°$. Overall, the results from the various gas models show little variation. Notably, the equilibrium gas model yields slightly larger values for both $X_{cp}$ and *PM*. This indicates that the effects of different models on the $X_{cp}$ and *PM* of the cone configuration are not significant, while those on control surfaces may be more pronounced, as suggested in [33].

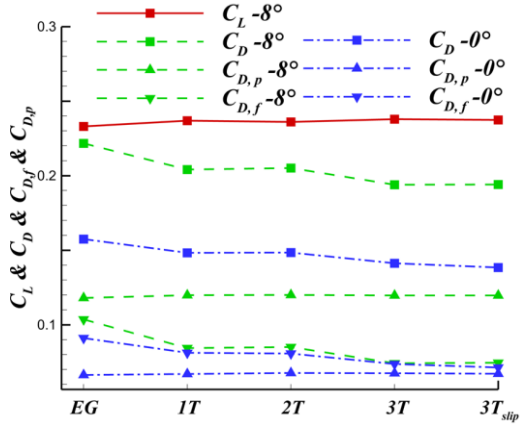
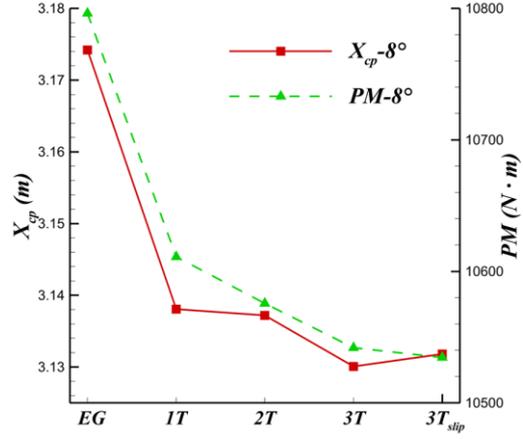

**Fig. 12 Aerodynamic characteristics predicted by different models at two AOAs**

**Fig. 13 Center-of-pressure location and pitching moment of the cone by different models at $\alpha = 8°$**

Fig. 14 presents $P_w$ distributions on the *SMP* at two AOAs, as predicted by different models. Fig. 14(a) shows that along the forebody, $P_w$ generally decreases initially and then remains relatively flat streamwisely. A sharp pressure drop occurs at the expansion corner due to intense flow expansion. The results by *EGM* except for being slightly lower near the nose, show little overall difference from the non-equilibrium results at both AOAs. Furthermore, the results from the different temperature models and the *3T$_{slip}$*-M all exhibit minimal variation. Fig. 14(b) shows that: (a) At $\alpha = 0°$, the base $P_w$ exhibits a symmetric distribution with a distinct maximum, while at $\alpha = 8°$ this maximum shifts leeward; (b) The base $P_w$ from the *EGM* is significantly lower than that from the non-equilibrium gases at both AOAs; (c) For the non-equilibrium models, the overall base $P_w$ is similar at the same $\alpha$, with the extrema at $\alpha = 0°$ satisfying $P_{w,2T} > P_{w,1T} > P_{w,3T} > P_{w,3T_{slip}}$, whereas at $\alpha = 8°$ the latter three values are approximately equal.

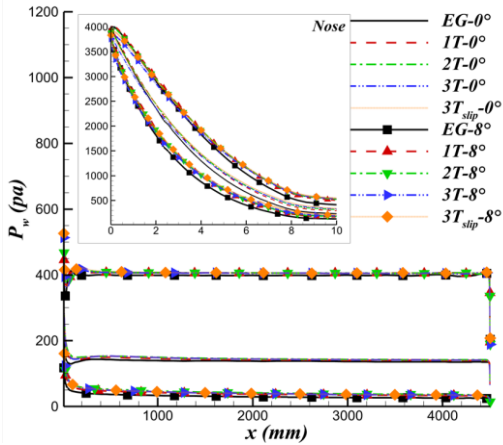
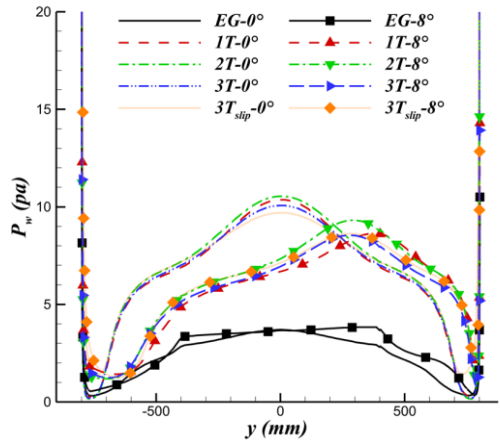

**(a) Forebody**          **(b) Base**

**Fig. 14 Wall pressure distributions on the *SMP* by different models at two AOAs**

Fig. 15 presents the skin friction coefficient ($C_f$) distributions on the cone *SMP* at two AOAs, as predicted by different models. The results show that: (a) At both AOAs, $C_f$ initially increases, then decreases, and finally levels off along the streamwise direction. At $\alpha = 8°$, the $C_f$ on the windward side is overall higher than that on the leeward side; (b) Except near the nose, the equilibrium gas yields a slightly higher $C_f$ than the non-equilibrium gases at $\alpha = 0°$. At $\alpha = 8°$, the difference is more pronounced on the windward side, while the values are similar on the leeward side; (c) For the non-equilibrium models near the nose at both AOAs, $C_f$ follows $C_{f,2T} > C_{f,1T} > C_{f,3T} > C_{f,3T_{slip}}$. At $\alpha = 8°$, $C_f$ along the windward side of the body, follows $C_{f,2T} \approx C_{f,1T} > C_{f,3T} \approx C_{f,3T_{slip}}$, while the values on the leeward side are approximately equal.

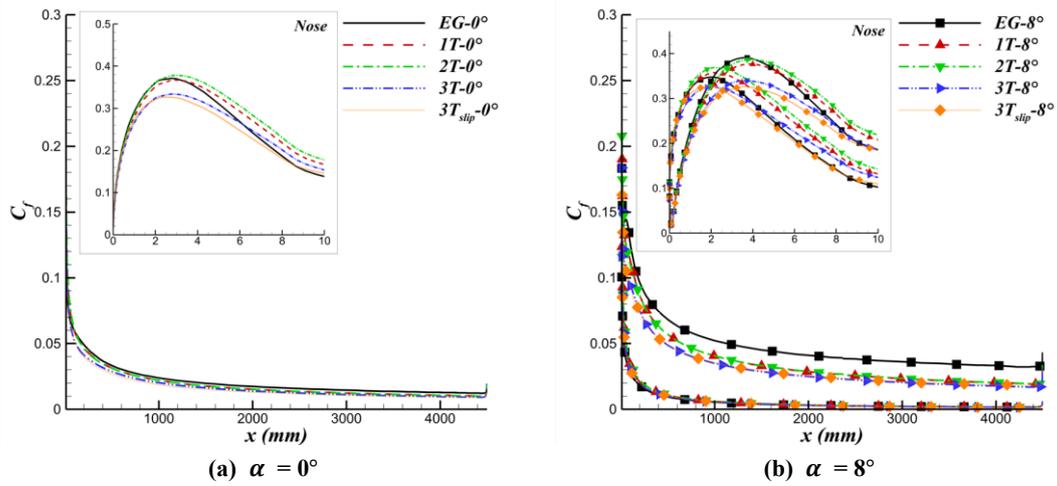

**(a) $\alpha = 0°$**          **(b) $\alpha = 8°$**

**Fig. 15 Skin friction distributions on forebody *SMP* by different models at two AOAs**

Fig. 16 and 17 present $Q_w$ distributions at two AOAs by different models on the forebody and base *SMP*. The results indicate that: (a) The $Q_w$ streamwisely along cone forebody decreases gradually at both AOAs. On the base, $Q_w$ exhibits a maximum—with the peak at the centerline for $\alpha = 0°$ whereas shifting leeward at $\alpha = 8°$. The findings are qualitatively consistent with base distribution of $Q_w$ shown in Fig. 10; (b) For the forebody at $\alpha = 0°$, the $Q_w$ by *EGM* is close to the non-equilibrium results, but the former are larger on both the windward and leeward sides at $\alpha = 8°$. On the base, the *EGM* predicts higher $Q_w$ at both AOAs; (c) Among the non-equilibrium models and near the nose, $Q_{w,2T} > Q_{w,1T} > Q_{w,3T} > Q_{w,3T_{slip}}$ at both AOAs, while along the forebody the differences are small. The peak of base $Q_w$ obeys $Q_{w,2T} \approx Q_{w,1T} > Q_{w,3T} > Q_{w,3T_{slip}}$ at $\alpha = 0°$, and follows $Q_{w,2T} > Q_{w,1T} \approx Q_{w,3T} > Q_{w,3T_{slip}}$ at $\alpha = 8°$. Overall, the $Q_w$ predicted by the *3T*-M is relatively lower.

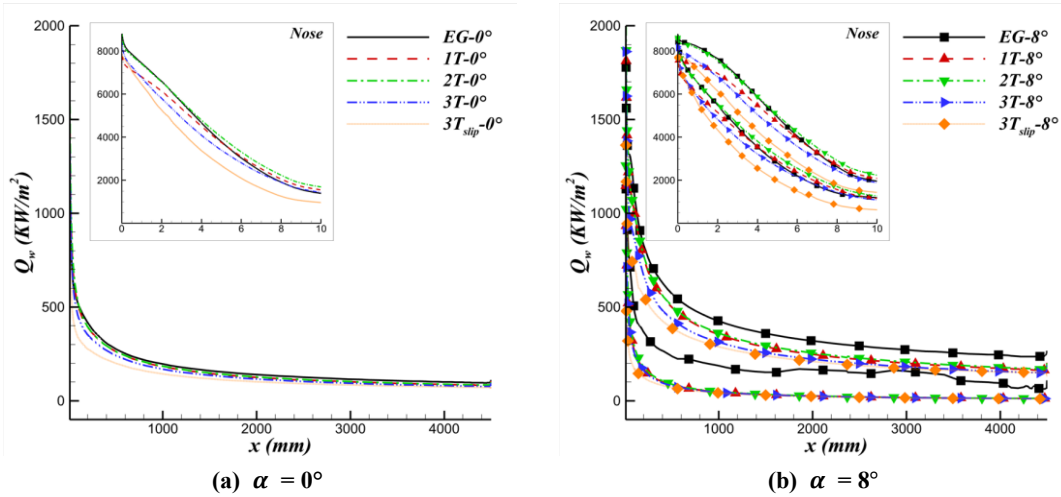

(a) $\alpha = 0°$  (b) $\alpha = 8°$

**Fig. 16 Wall heat flux distributions on the forebody *SMP* by different models at two AOAs**

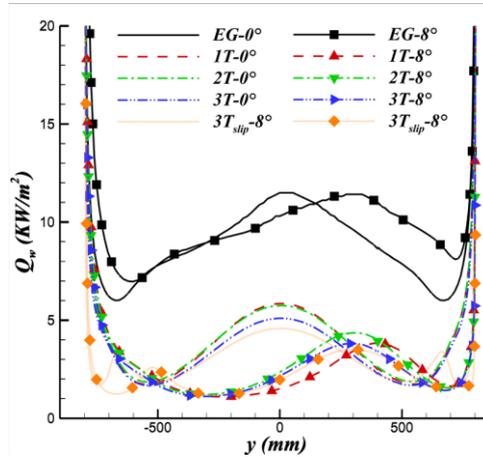

**Fig. 17 Wall heat flux distributions on the *SMP* of the base by different models at two AOAs**

3.3 Comparative analysis of equilibrium and non-equilibrium predictions

To compare the equilibrium and non-equilibrium characteristics predicted by different models, the cone flow is examined from several perspectives: the nose, body, and base regions. A qualitative and quantitative analysis is conducted on key characteristics, e.g., the specific heat ratio ($\gamma$), temperatures, and species mass fractions at both AOAs. The overall differences in these predictions by the various models are analyzed first.

3.3.1 Overall analysis of differences in equilibrium and non-equilibrium predictions

(1) Case at $\alpha = 0°$

Since the flow field is axisymmetric at $\alpha = 0°$, Fig. 18 concisely presents the differences in $\gamma$ on the cone horizontal plane ($y = 0$) as predicted by the various models. An inset provides a close-up view of the nose region, and the solid white line denotes the difference the is zero (similar cases hereafter will not be reiterated). The comparison reveals that the differences in $\gamma$ among the models are primarily concentrated within the shock layer and the wake field. Specifically, the overall values are all less than 1.4. This reduction stems from the excitation of internal degrees of freedom within the gas molecules, primarily vibrational energy. This excitation increases both the specific heat capacity at constant volume ($C_v$) and that at constant pressure ($C_p$), while keeping their difference constant, thus lowering $\gamma$ (where $\gamma = C_p/C_v$). In the regions near the nose and along the base centerline, the $\gamma$ follows $\gamma_{EG} < \gamma_{1T}$. This occurs because the *EGM* allocates a portion of the energy to be stored in the form of chemical potential, rather than being fully manifested as a temperature rise. This allocation leads to an increase in the specific heat capacity, thereby further reducing $\gamma$. Among the different temperature models, $\gamma_{1T}$ is generally lower than $\gamma_{2T}$ and $\gamma_{3T}$, while the latter two are close in value. This is because the *1T*-M assumes equipartition of energy among translational-rotational, vibrational, and electronic modes. In contrast, the *2T*- and *3T*-M account for different energy transfer rates, where the vibrational and electronic modes have longer relaxation times and thus a lower energy fraction. This results in relatively lower values for $T_{ve/v}$, a correspondingly smaller $C_v$, and consequently a higher $\gamma$.

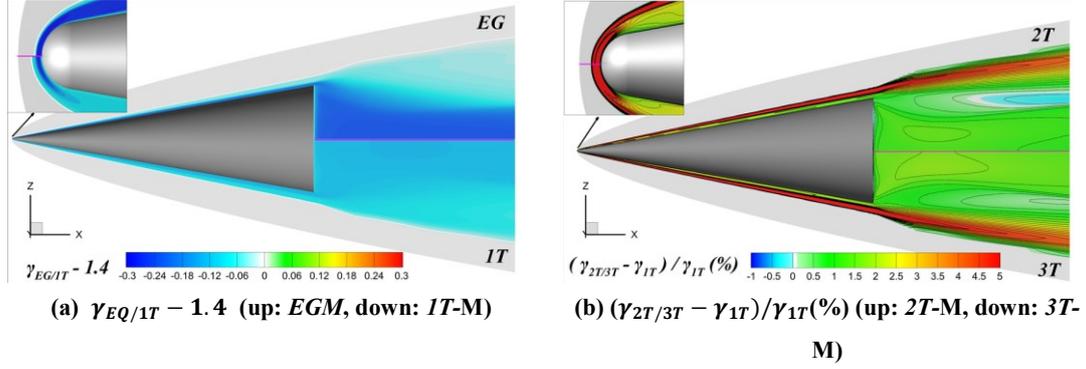

(a) $\gamma_{EQ/1T} - 1.4$ (up: *EGM*, down: *1T*-M)  (b) $(\gamma_{2T/3T} - \gamma_{1T})/\gamma_{1T}(\%)$ (up: *2T*-M, down: *3T*-M)

**Fig. 18 Distributions of the specific heat ratio difference on the horizontal plane ($y = 0$) at $\alpha = 0°$, as predicted by different models, where the approximately white region denotes the zero-difference**

Fig. 19 presents the distributions of temperature differences on the horizontal plane ($y = 0$) at $\alpha = 0°$, as predicted by different models. It can be seen that the discrepancies are similarly concentrated in the shock layer and the wake field. Fig. 19(a) shows that except in the vicinity of the base expansion region and localized areas near the wall (shown in blue), $T_{EG}$ is significantly lower than $T_{1T}$, with the temperature difference in the base flow field reaching approximately 3000 K. This is primarily because, in the *EGM*, a substantial portion of the internal energy is converted into chemical potential energy, resulting in a relatively lower temperature. In the expansion region, however, the chemical non-equilibrium in the *1T*-M yields a "freezing" of recombination and other reactions. The energy required for expansion work in this region is mainly drawn from the internal energy, causing $T_{1T}$ to decrease more sharply than $T_{EG}$ does, which makes $T_{EG}$ slightly higher locally. It is noteworthy that the occurrence is located in the downstream area extending along the conical surface, rather than the region around base shoulder. Fig. 19(a) also shows that excluding the vicinity of the expansion region (shown in blue), $T_{1T} < T_{tr,2T}$ in the base flow with a maximum difference of about 1200 K. This stems from their different energy partitioning. In the *1T*-M, equipartition of energy is assumed between translational-rotational and vibrational-electronic modes. In contrast, the *2T*-M accounts for different energy exchange between modes. The translational-rotational mode, having a higher energy exchange rate, retains a larger share of the total energy, which consequently results in a higher $T_{tr,2T}$. The lower $T_{tr,2T}$ in the expansion region is due to a reduced molecular collision frequency, which causes the vibrational energy to become "frozen". Consequently, the energy required for the expansion work is drawn primarily from the translational-rotational energy, leading to a more pronounced decrease in $T_{tr,2T}$ due to the expansion.

Fig. 19(b) shows that except for localized areas near the nose (shown in blue), the distributions satisfy $T_{tr,2T} > T_{tr,3T}$ and $T_{ve,2T} > T_{v,3T}$. The former relationship arises because the decoupling of vibrational and electronic modes as well as a lower $T_e$ in the *3T*-M leads to a lower $T_c$ for the reverse reactions of Reactions 7–9 in Table 1. This results in a weakening of the associated exothermic reactions. Consequently, compared to the *2T*-M, the conversion of chemical energy into translational-rotational energy is suppressed, leading to a lower $T_{tr,3T}$. The latter relationship is because the *2T*-M combines the vibrational and electronic degrees of freedom into a single description. This amalgamation impedes the efficient energy transfer between electronic and translational-rotational modes, causing energy to be retained within this combined mode and resulting in a relatively higher $T_{ve,2T}$. Besides, it should be noted that the blue region near the nose is primarily attributed to the different starting positions of the shock waves in the two models (see Fig. 24(a)), leading to a temperature discrepancy at the same spatial location. This difference is not directly attributed to the physical mechanisms discussed above.

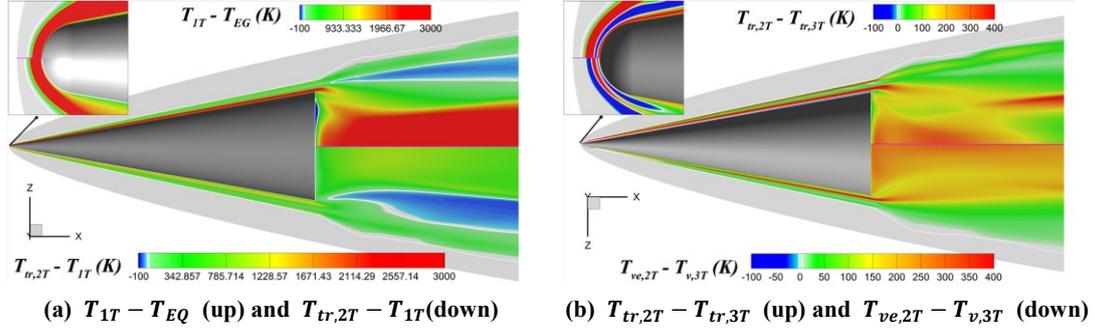

(a) $T_{1T} - T_{EQ}$ (up) and $T_{tr,2T} - T_{1T}$ (down)  (b) $T_{tr,2T} - T_{tr,3T}$ (up) and $T_{ve,2T} - T_{v,3T}$ (down)

Fig. 19 Distributions of temperature differences on the horizontal plane ($y = 0$) at $\alpha = 0°$, as predicted by different models, where the approximately white region denotes the zero-difference

To investigate the prediction differences regarding the thermal non-equilibrium extent between *2T*- and *3T*-M, Fig. 20 presents the distributions of $T_{tr}/T_{ve/v}$ and the fraction of vibrational-electronic/vibrational energy ($E_{ve/v}$) in the specific internal energy ($E - 0.5U^2$) on the horizontal plane ($y = 0$) at $\alpha = 0°$. The regions approximately in white indicate the value therein be near 1 (similar cases hereafter will not be repeatedly explained). The figure reveals that the thermally non-equilibrium regions are primarily concentrated in the post-shock flow near the nose and body, as well as in parts of the wake. In detail, Fig. 20(a) shows that the distributions of $T_{tr,2T}/T_{ve,2T}$ and $T_{tr,3T}/T_{v,3T}$ are similar—except in the region between the *EW* and the *RCW* at the base (shown in blue), the relationship $T_{tr} > T_{ve/v}$ generally holds. This is due to the relatively long vibrational relaxation time, resulting in a smaller fraction of $E_{ve/v}$ and consequently lower $T_{ve/v}$. After passing through the *EW*, $T_{tr}$ decreases rapidly. However, because of its longer relaxation time and potential "freezing" across the *EW*, $T_{ve/v}$ does not decrease as much, leading to $T_{tr} < T_{ve/v}$ within the blue region. Fig. 20(b) shows that the proportion distributions of $E_{ve,2T}$ and $E_{v,3T}$ to the internal energy are roughly similar, mainly concentrated near the wall and in the base flow region. The former is larger in the vicinity of the base wall and along the downstream centerline. This difference stems from their distinct energy exchange pathways, the specifics of which align with the reasons for $T_{ve,2T} > T_{v,3T}$ as previously discussed.

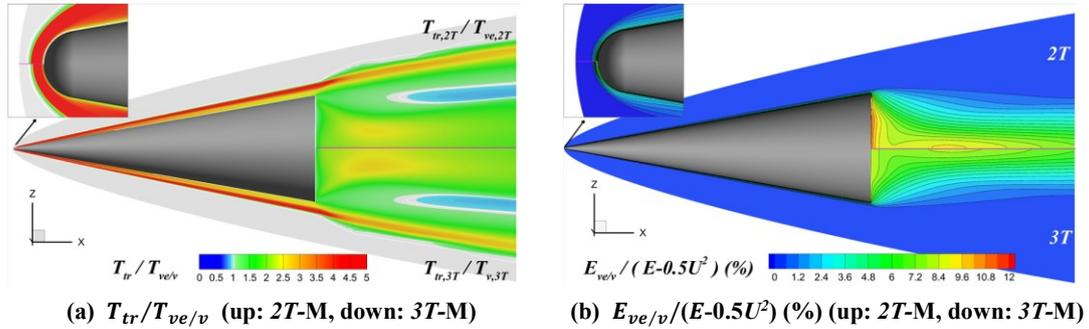

(a) $T_{tr}/T_{ve/v}$ (up: *2T*-M, down: *3T*-M)  (b) $E_{ve/v}/(E-0.5U^2)$ (%) (up: *2T*-M, down: *3T*-M)

Fig. 20 Comparison of $T_{tr}/T_{ve/v}$ and $E_{ve/v}/(E-0.5U^2)$ (%) distributions on the horizontal plane ($y = 0$) by

*2T-* and *3T*-M at $\alpha = 0°$, where the approximately white region denotes the ratio being near 1

(2) Case at $\alpha = 8°$

At $\alpha = 8°$, the differences in predictions by the various models closely resemble those at $\alpha = 0°$. The non-equilibrium characteristics of the flow field remain concentrated in the shock layer and wake region, but their specific distributions and intensity are altered by $\alpha$. Consequently, only selected comparisons of the differences in flow field quantities from some models are presented for this condition. Fig. 21 presents the distributions of differences of $\gamma$ and temperature on the *SMP* and streamwise cross-sections at $\alpha = 8°$. The results in Fig. 21(a) and (b) show that the regions of maximum difference for both $\gamma_{1T} - \gamma_{EG}$ and $T_{1T} - T_{EG}$ at the base are shifted leeward due to $\alpha$. Within the base flow region, $\gamma_{1T} > \gamma_{EG}$ (with the maximum difference reaching approximately 0.12), for reasons consistent with the $\alpha = 0°$ case. In the region outward from this area to the shock, $\gamma_{1T} < \gamma_{EG}$, and the biggest difference occurs there because the shock standoff distance by the *1T*-M is greater than that by the *EGM* (see Fig. 24(b)). Consequently, $\gamma_{1T}$ begins to decrease while $\gamma_{EG}$ remains unchanged, creating this disparity. Fig. 21(b) shows that $T_{1T} < T_{EG}$ in the base region and in a semi-annular expansion zone on the downstream windward side. Conversely, the opposite relationship holds in the region outward from these areas to the shock, with the maximum difference at the base reaching approximately 3000 K.

The results in Fig. 21(c) show that $T_{tr,2T} > T_{1T}$ holds except in some regions affected by expansion (shown in blue), with a maximum difference at the base of approximately 1200 K. Fig. 21(d) indicates that $T_{ve,2T} > T_{v,3T}$ except in a localized area of the shock layer near the nose, where the maximum difference at the base is about 500 K. The distribution near the nose is primarily due to the variation in shock position among the different models, which amplifies the differences in the temperature distributions. Since the underlying causes of the temperature differences among models are similar to those at $\alpha = 0°$, they are not repeated here.

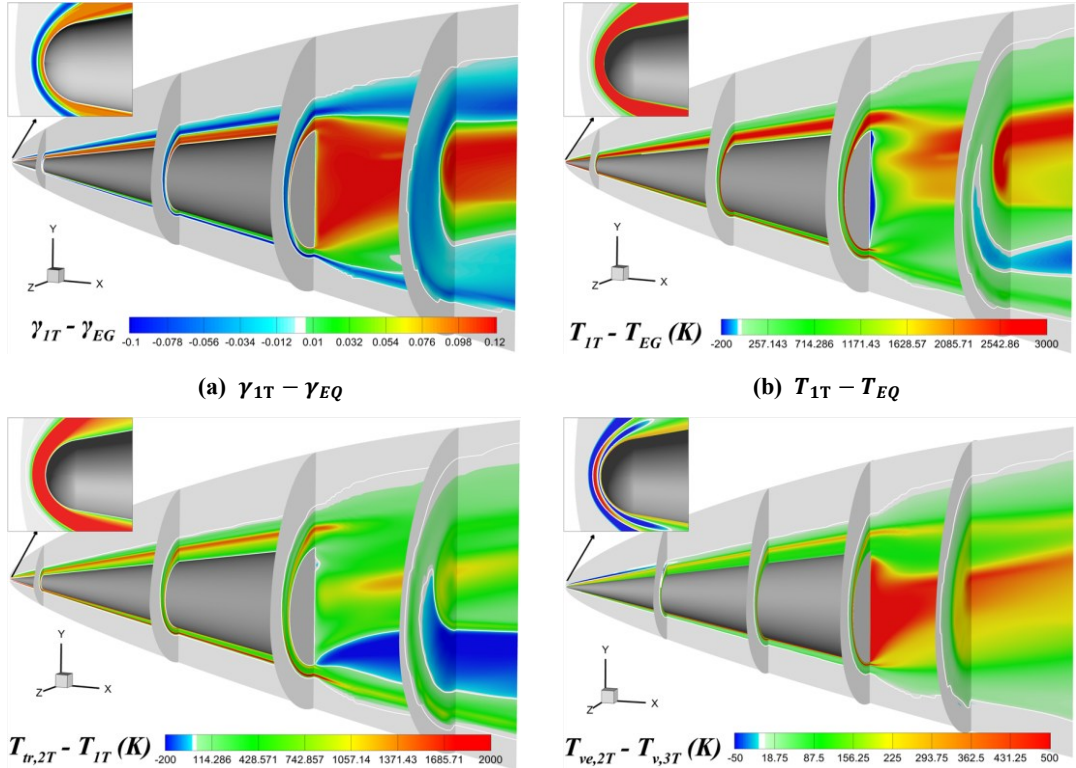

(a) $\gamma_{1T} - \gamma_{EQ}$

(b) $T_{1T} - T_{EQ}$

(c) $T_{tr,2T} - T_{1T}$  (d) $T_{ve,2T} - T_{v,3T}$

**Fig. 21 Distributions of differences of quantities on the *SMP* and streamwise cross-sections by different models at $\alpha = 8°$, where the approximately white region indicates the zero-difference**

Taking the *2T*-M as an example, Fig. 22 presents the distribution of $T_{tr}/T_{ve}$ on the *SMP* and streamwise cross-sections at $\alpha = 8°$. It can be seen that, except in some regions affected by expansion (shown in blue), $T_{tr,2T} > T_{ve,2T}$ holds, for reasons explained in analysis of the case $\alpha=0°$. In the expansion regions, $T_{tr}$ decreases rapidly, while $T_{ve}$ changes more slowly because of the longer vibrational-electronic energy relaxation, which leads to $T_{tr,2T} < T_{ve,2T}$ thereby.

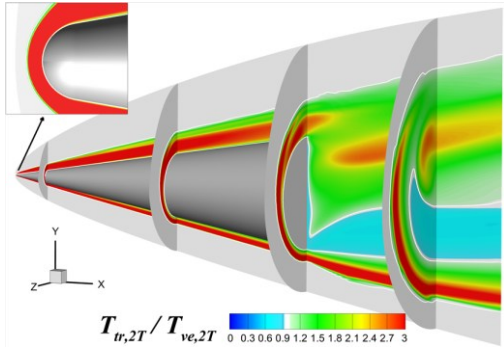
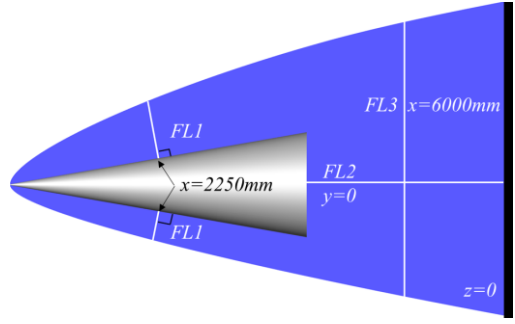

**Fig. 22 Distributions of $T_{tr,2T}/T_{ve,2T}$ on the *SMP* and streamwise cross-sections by *2T*-M at $\alpha = 8°$, where the approximately white region denote the ratio being near 1**

**Fig. 23 Schematic diagram of feature lines for quantitative distribution analysis**

3.3.2 Comparative analysis of quantitative distributions along featured lines at forebody

To reveal the differences in cone flow characteristics predicted by various models, quantitative analyses are conducted in this section on key quantities, e.g., temperatures, $\gamma$, and $Y_i$, along specified feature lines (*FL*). The locations of selected *FL*s are illustrated in Fig. 23. Specifically, *FL1* is a wall-normal line originated form the wall at $x = 2250$ mm on the *SMP*, which splits into windward and leeward cases when $\alpha = 8°$; *FL2* is the centerline at the base ($y, z = 0$); and *FL3* lies on the *SMP* at $x = 6000$ mm and is perpendicular to *FL2*. This subsection focuses solely on the stagnation line at the nose and the *FL1* location; *FL2* and *FL3* will be detailed in Section 3.3.3. The following first introduces the distributions of quantities along the cone stagnation line at $\alpha = 0°$.

(1) The nose section

A comparison of the temperature, $\gamma$, and $Y_i$ distributions along the stagnation line at $\alpha = 0°$, as predicted by different models, is presented in Fig. 24. The results in Fig. 24(a) show that with increase of $x$: (a) $T_{EG}$, $T_{1T}$, $T_{tr,2T}$, and $T_{tr,3T}$ all increase after passing through the shock and exhibit the typical peak profile within the temperature boundary layer. It is noteworthy that due to the quite small nose radius, the flow is featured by a low $Re$ characteristic, leading to temperature distributions different from those in Fig. 2(a). Among the different models, the temperature peaks follow $T_{EG} < T_{1T} < T_{tr,2T} < T_{tr,3T}$; (b) After the gas passes through the shock, vibrational energy begins to be excited; $T_{ve,2T}$ and $T_{v,3T}$ exhibit distributions consistent with the canonical temperature boundary layer but with lower magnitudes, showing distinct non-equilibrium, and $T_{ve,2T}$ is slightly greater than $T_{v,3T}$; (c) Electronic energy also increases after the shock. With increasing $x$, the $T_e$ distribution shows a more gradual rise, and its profile in the near-wall region is shaped by the constraints imposed by the boundary conditions (see Section 2.3). (d) The trends of distributions of $T_{tr,3T_{slip}}$ and $T_{e,3T_{slip}}$ are similar to their non-slip results. However, influenced by the boundary condition, $T_{v,3T_{slip}}$ exhibits a near-adiabatic distribution when approaching the wall, forming a significant contrast with that of $T_{v,3T}$ under the isothermal wall condition.

Fig. 24(b) shows that as $x$ increases, except for $\gamma_{3T_{slip}}$ which assumes a zero-normal-gradient distribution after decreasing, the $\gamma$ by other models generally first decrease, reach a minimum, and then increase. The minimum $\gamma$ values satisfy $\gamma_{EG} < \gamma_{1T} < (\gamma_{2T} \approx \gamma_{3T})$, with specific reasons detailed in Section 3.3.1. Fig. 24(c) presents the distributions of $Y_N$, $Y_{NO}$, and $Y_{NO^+}$ along $x$, with an inset showing the $Y_i$ distributions on a logarithmic scale to magnify the differences near the wall. It can be observed that the magnitudes of $Y_{NO^+}$ are very small for all models. Near the wall, both $Y_N$ and $Y_{NO}$ satisfy $Y_{1T} > Y_{2/3T}$. This is because the $T_c$ in the *1T*-M (see Eq. (6)) directly employs a higher single temperature, whereas the *2T*- or *3T*-M use two or three temperatures to compute $T_c$ (see Eqs. (10) and (11)). Since $T_{ve/v}$ and $T_e$ are relatively low under this computational condition, the $T_c$ in the latter models is smaller, leading to reduced reaction rates and consequently lower $Y_i$. Furthermore, the results by the *3T$_{slip}$*-M show a decreasing trend for all $Y_i$ near the wall, which may be due to the nonzero slip velocity at the wall and consequent reduction of chemical reaction extent.

In short, the distributions of various predictions along the stagnation line have been analyzed on behalf of nose flow. It is noteworthy that given the small nose radius, its overall impact on the whole aerodynamic characteristics is limited.

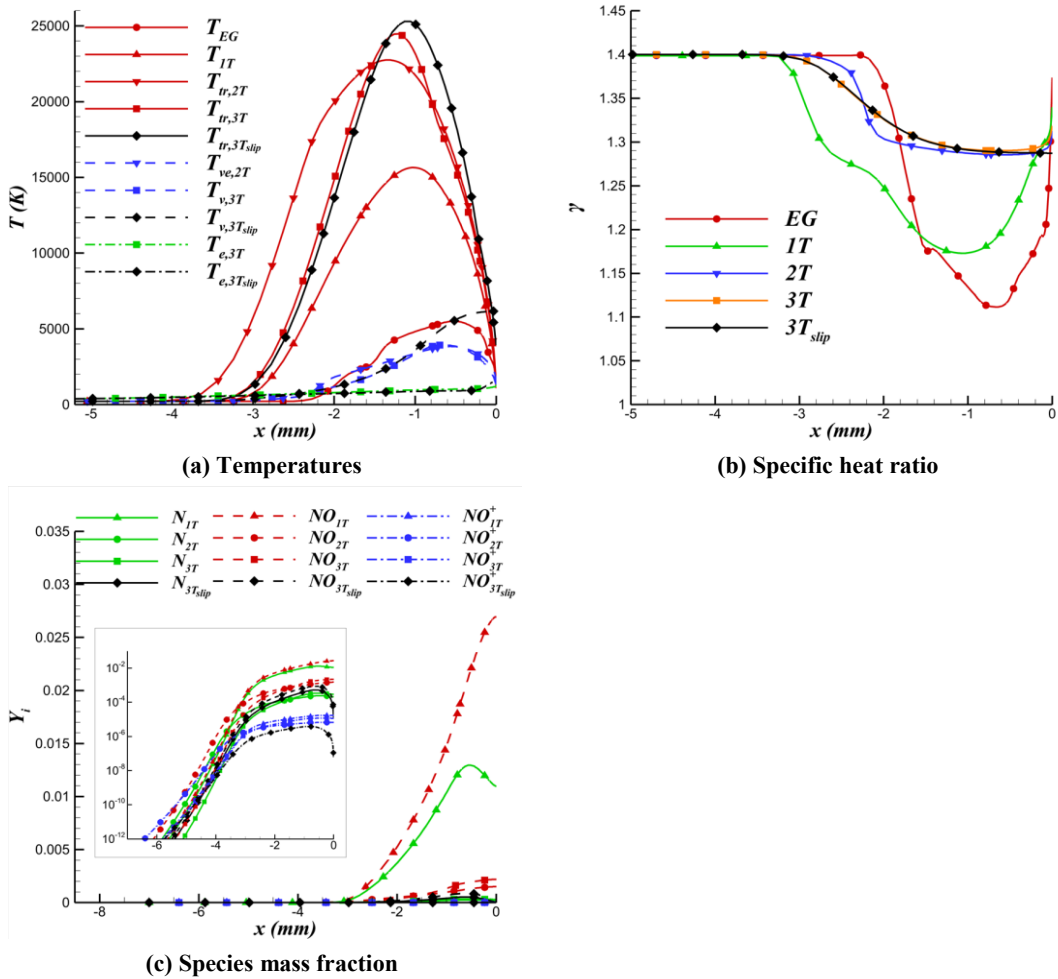

(a) Temperatures

(b) Specific heat ratio

(c) Species mass fraction

**Fig. 24 Distributions of quantities along the stagnation line at the nose by different models at $\alpha = 0°$**

(2) The body section

Fig. 25 first presents the distributions, e.g., those of temperature, $\gamma$, velocity, and $Y_i$, along *FL1* (wall-normal direction) at $\alpha = 0°$, as predicted by different models. In the figure, $y_n$ denotes the coordinate along the wall-normal direction, and $u_t$ represents the velocity component tangential to the wall. For clarity, an inset provides a close-up view of the near-wall $u_t$ and $Y_i$ distributions, with the latter shown on a logarithmic scale.

Fig. 25(a) shows that along with decrease of $y_n$: (a) After the freestream passes through the shock, the post-shock $T_{EG}$, $T_{1T}$, $T_{tr,2T}$, and $T_{tr,3T}$ increase sharply, following the shock relations, and then exhibit an approximate plateau (qualitatively consistent with inviscid conical flow characteristics). Subsequently, a temperature peak appears within the boundary layer, and the peak values satisfy $T_{EG} < T_{1T} < T_{tr,2T} \approx T_{tr,3T}$. The primary reasons of the relationship have been given in explanation of Fig. 19 in Section 3.3.1; (b) After the gas passes through the shock, vibrational energy begins to be excited. The distributions of $T_{ve,2T}$ and $T_{v,3T}$ gradually increase as $y_n$ decreases, forming profiles consistent with those of $T_{tr}$ but at lower magnitudes, thus demonstrating pronounced non-equilibrium characteristics. Their peaks satisfy $T_{ve,2T} > T_{v,3T}$, which aligns with the size relationship at the corresponding location in Fig. 19(b); (c) Electronic energy also increases after the shock. As $y_n$ decreases, the $T_e$ distribution exhibits a more gradual increase. The profile of $T_e$ in the near-wall region is shaped by the constraints of the boundary conditions (see Section 2.3); (d) Regarding the differences between slip and non-slip results, the distributions of $T_{tr,3T_{slip}}$ and $T_{e,3T_{slip}}$ are similar to their non-slip counterparts. However, influenced by the boundary condition, $T_{v,3T_{slip}}$ exhibits a near-adiabatic distribution when approaching the wall, forming a significant contrast with the $T_{v,3T}$ distribution under the isothermal wall condition.

Fig. 25(b) shows that as $y_n$ decreases, except for $\gamma_{3T_{slip}}$ which indicates a zero-normal-gradient distribution, the $\gamma$ from all other models generally first decrease, reach a minimum, and then increase. Among them, the variation patterns of $\gamma_{EG}$ and $\gamma_{1T}$ correspond to those of $T_{EG}$ and $T_{1T}$, such as similar platform regions. Similarly, the patterns of $\gamma_{2T}$, $\gamma_{3T}$, and $\gamma_{3T_{slip}}$ show somewhat correspondences with $T_{ve,2T}$, $T_{v,3T}$, and $T_{v,3T_{slip}}$, respectively. The reasons are as follows. For *EGM*, where energy modes are equally partitioned and share a single temperature, $C_p$ is calculated using fitting formulas as a function of $T$. For the *3T*-M, $C_v = C_{v,tr} + C_{v,v} + C_{v,e}$ where $C_{v,tr}$ depends only on the species; $C_{v,v} = \partial e_v/\partial T_v$, and $C_{v,e} = \partial e_e/\partial T_e$. Given the extremely low concentration of electronic species and the low $T_e$ under this condition, $\gamma$ is predominantly influenced by $T_v$. In the *2T*-M, where $T_v = T_e = T_{ve}$, the variation in $\gamma$ is thus related to $T_{ve}$. Because $C_p$ possesses the constant difference with respect to $C_v$, the former has the similar relationship with corresponding temperature, and therefore the distributions of $\gamma$ can be related to the dependent temperatures.

Fig. 25(c) presents the wall-normal distribution of the $u_t$. The results show that the $u_t$ profile by the *EGM* features an overall thinner boundary layer and stronger shear compared to the non-equilibrium models. This is consistent with its temperature distribution in Fig. 25(a), where the relatively lower temperature and internal energy lead to higher kinetic energy and velocity. The enlarged inset reveals that under the non-slip boundary condition, $u_{t,3T}$ is the lowest near the wall, $u_{t,EG}$ is the highest, and the others lie in between, which aligns with the $C_D$ results in Fig. 12. Furthermore, $u_{t,3T_{slip}}$ at the wall is approximately 200 m/s, but the results indicate that slip does not significantly alter the velocity shear magnitude (see Fig. 15). It is noteworthy that such understanding be significant for engineering because skin friction is extremely concerned in vehicle design. Fig. 25(d) shows the wall-normal distributions of $Y_N$, $Y_{NO}$, and $Y_{NO^+}$. These species are primarily confined within the boundary layer. The magnitudes of $Y_{NO^+}$ are small for all models, while those of $Y_N$ and $Y_{NO}$ by the *1T*-M are relatively larger.

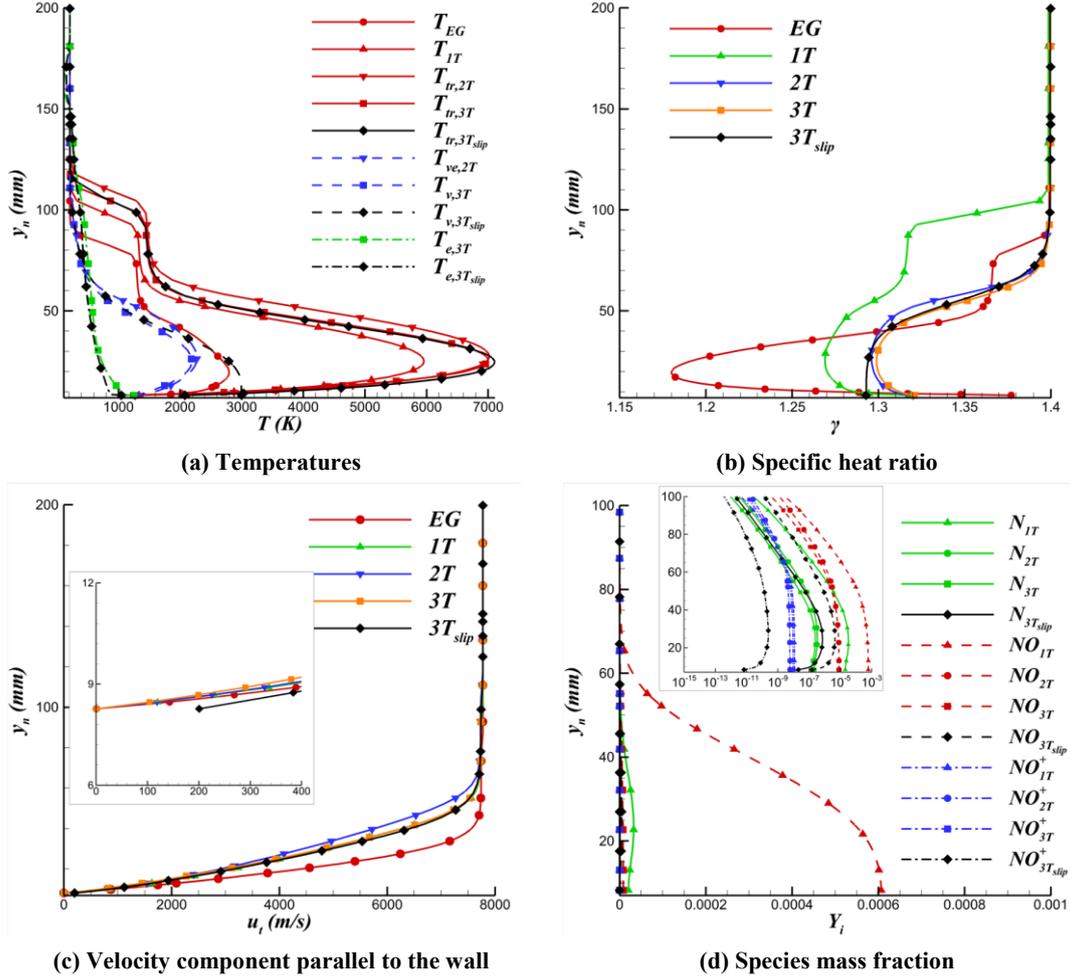

(a) Temperatures

(b) Specific heat ratio

(c) Velocity component parallel to the wall

(d) Species mass fraction

Fig. 25 Distributions of quantities along *FL1* at body section by different models at $\alpha = 0°$

At $\alpha = 8°$, the distributions of temperatures, $\gamma$, $u_t$, and $Y_i$ along *FL1* (comprising windward and leeward one respectively) predicted by the different models are overall similar to those at $\alpha = 0°$ (detailed results are provided in the Appendix). Therefore, they are not discussed in detail here. Some obvious differences are as follows: On the windward side, the stronger shock leads to a reduced boundary layer thickness and a more pronounced platform-like distribution in $u_t$. Conversely, on the leeward side, the weaker shock results in a relatively thicker boundary layer, and the platform-like regions observed in $T$, $T_{tr}$, $\gamma_{EG}$, and $\gamma_{1T}$ disappear.

3.3.3 Comparison and analysis of base flow predictions and distributions on feature lines

As noted in the introduction, the structure and physicochemical characteristics of the base flow are of significant importance for engineering. To elucidate and compare the characteristics of the base flow by the various models, qualitative and quantitative analyses are conducted on the base flow and the distributions of temperatures, $\gamma$, and $Y_i$ along *FL*s. The results for the case at $\alpha = 0°$ are discussed first.

(1) Case at $\alpha=0°$

(a) Temperatures and $\gamma$

Fig. 26 first presents a comparison of the temperature distributions on the horizontal plane ($y = 0$) at the base at $\alpha = 0°$. Since the structural similarity of $T$ and $T_{tr}$ by different models, only the $T_{1T}$ distribution is used as an example for comparative analysis with $T_{EG}$. The results in Fig. 26(a) show that both $T_{EG}$ and $T_{1T}$ exhibit a high-temperature center structure at the base. Combined with Fig. 6, the formation of this structure can be attributed to the flow separation and subsequent recirculation zone (with flow stagnation downstream), which converts part of the kinetic energy of the upstream boundary layer into internal energy, thereby increasing the temperature. At this high-temperature center, $T_{EG} < T_{1T}$, which is consistent with the region where $T_{1T} - T_{EG}$ shows the maximum difference in Fig. 19(a). Fig. 26(b) presents a comparison of the $T_{ve,2T}$ and $T_{v,3T}$ distributions at the cone base. The results show that their distribution structures are similar, with higher temperatures appearing near the downstream centerline of the base, where $T_{ve,2T}$ exhibits relatively larger values. Fig. 26(c) and (d) further compare the $T_v$ and $T_e$ distributions by the *3T*- and *3T$_{slip}$*-M. It can be observed that $T_{v,3T_{slip}}$ exhibits a more uniform and significantly higher distribution near the base wall compared to $T_{v,3T}$, due to the temperature jump effect. $T_e$ is primarily and uniformly distributed within the base and downstream regions under both slip and non-slip conditions, with $T_{e,3T} > T_{e,3T_{slip}}$ observed near the base and along the downstream centerline.

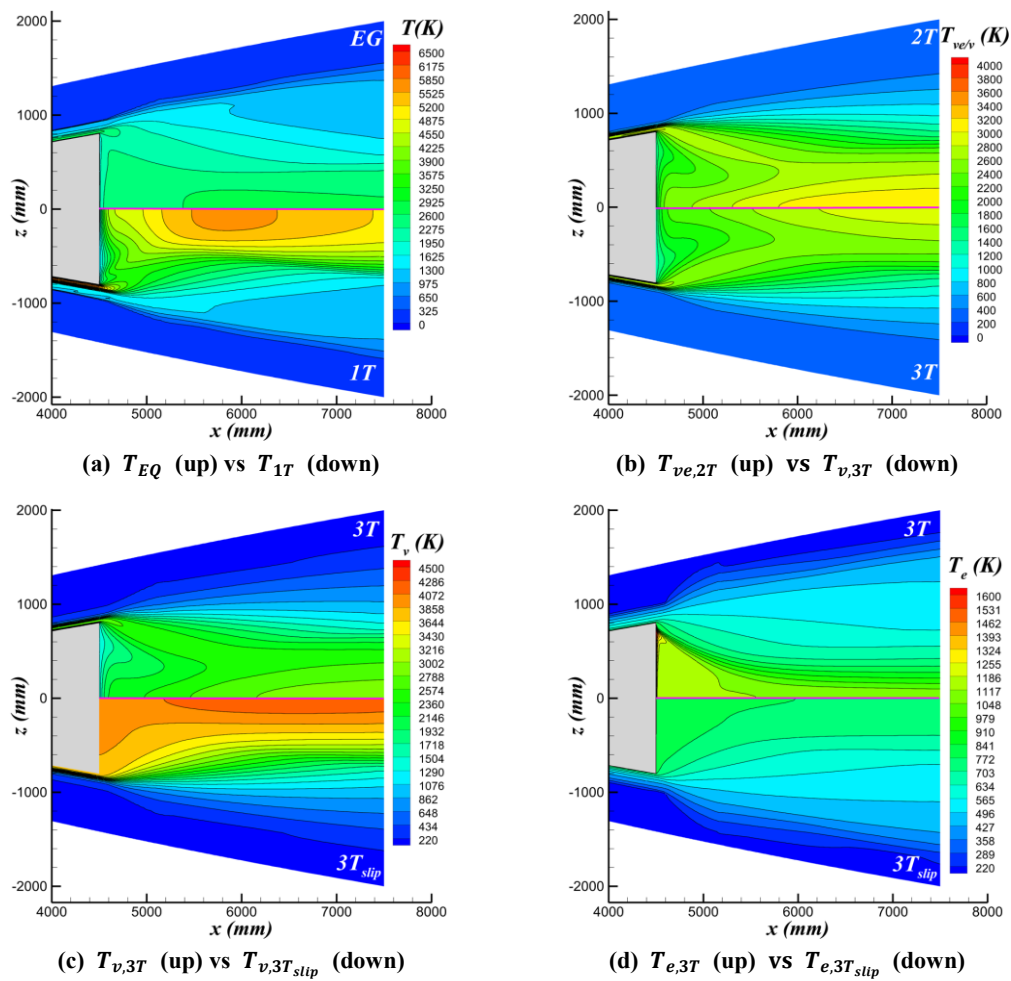

(a) $T_{EQ}$ (up) vs $T_{1T}$ (down)  
(b) $T_{ve,2T}$ (up) vs $T_{v,3T}$ (down)  
(c) $T_{v,3T}$ (up) vs $T_{v,3T_{slip}}$ (down)  
(d) $T_{e,3T}$ (up) vs $T_{e,3T_{slip}}$ (down)  

Fig. 26 Comparison of temperature distributions on horizontal plane ($y = 0$) at the base by different models at $\alpha = 0°$

To quantify the differences in $T$ and $T_{tr}$ distributions among the various models, the temperature extrema ($T_{EXT}$) at the base flow and their streamwise locations, $x_{T_{EXT}}$, at $\alpha = 0°$ were selected as metrics. The corresponding values are extracted and presented in Fig. 27. Given the axial symmetry at $\alpha = 0°$, the extrema are located on the base centerline ($y, z = 0$). The results show that compared to the *1T*-M, the *EGM* yields a smaller $T_{EXT}$ but a larger $x_{T_{EXT}}$, which is consistent with the qualitative findings in Fig. 26(a). Among the different non-equilibrium models, the $x_{T_{EXT}}$ values are similar, while the magnitudes satisfy $T_{EXT,1T} < T_{tr_{EXT},3T} < T_{tr_{EXT},2T}$. For the *3T$_{slip}$*-M, its $T_{EXT}$ is greater than that of the non-slip case, whereas the $x_{T_{EXT}}$ shows little difference.

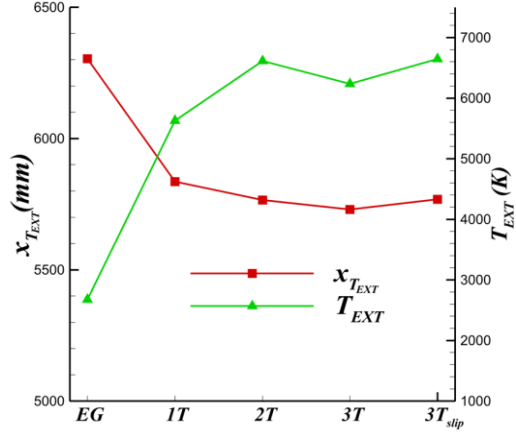

**Fig. 27 Extrema of $T$ and $T_{tr}$ as well as their streamwise locations by different models at $\alpha = 0°$**

To further quantify the influence of different models on the quantities within the cone base flow field at $\alpha = 0°$, Fig. 28 and 29 present comparative distributions of temperature and $\gamma$ along *FL2* and *FL3*, respectively. The specific locations of *FL2* and *FL3* are defined in Fig. 23. Fig. 28(a) shows the following trends along *FL2* with increasing $x$: (a) $T_{EG}$, $T_{1T}$, $T_{tr,2T}$, and $T_{tr,3T}$ increase rapidly near the base, reach a maximum at $x \approx 5700$ mm, and then decrease slowly. The magnitude and location of this maximum correspond to those in Fig. 27. The temperatures generally satisfy $T_{EG} \ll T_{1T} < T_{tr,3T} < T_{tr,2T}$ across the models, consistent with the results in Fig. 19. The specific reasons are detailed in Section 3.3.1; (b) $T_{ve,2T}$ and $T_{v,3T}$ primarily exhibit an initial increase followed by a slow variation, with $T_{ve,2T} > T_{v,3T}$, aligning with the findings in Fig. 19(b). Furthermore, the $T_{ve/v}$ predicted by the *2T*- and *3T*-M are lower than their corresponding $T_{tr}$, demonstrating clear non-equilibrium characteristics. $T_e$ near the base follows exhibits a distribution constrained by the boundary conditions, while downstream it remains essentially constant; (c) The distribution trends of $T_{tr,3T_{slip}}$ and $T_{e,3T_{slip}}$ are similar to their non-slip counterparts. Notably, due to the vibrational energy jump condition (see Eq. (14)), $T_{v,3T_{slip}}$ exhibits a distribution significantly higher than that of $T_{v,3T}$ (and also $T_{ve,2T}$). Fig. 28(b) indicates that, except for $\gamma_{3T_{slip}}$ which shows a zero-normal-gradient distribution near the base wall, the $\gamma$ of other models first decrease rapidly and then vary gradually along *FL2*. Among the different models, the $\gamma$ values generally satisfy $\gamma_{EG} < \gamma_{1T} < (\gamma_{2T} \approx \gamma_{3T})$, in agreement with the results in Fig. 18.

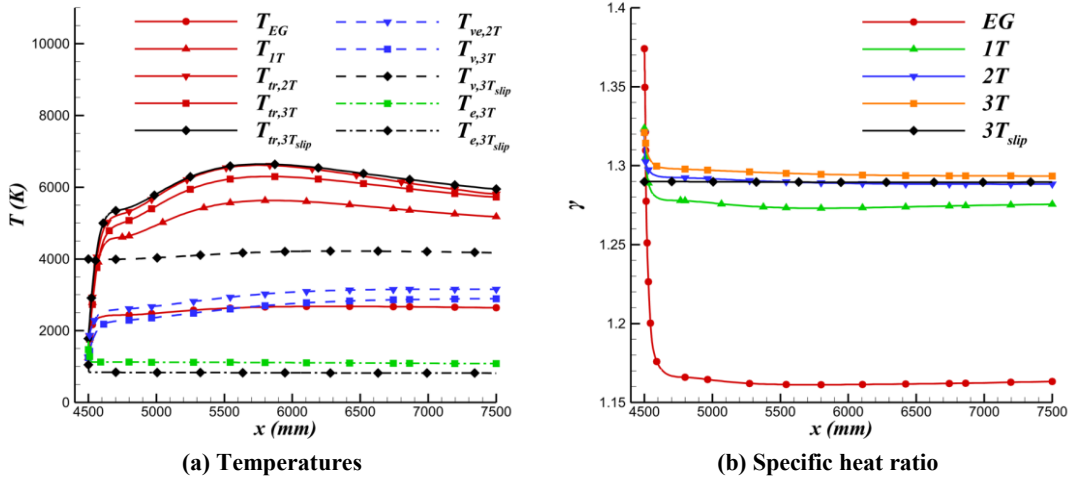

(a) Temperatures  (b) Specific heat ratio

**Fig. 28** Distributions of temperatures and $\gamma$ along *FL2* versus $x$ by different models at $\alpha = 0°$

The relationships of temperatures and $\gamma$ from different models along *FL3* are consistent with those along *FL2*. Therefore, the following discussion focuses on their distribution patterns there. The results in Fig. 29 show that the distributions along *FL3* show symmetry about $y = 0$ as expected. As $y$ decreases to zero: (a) After passing through the shock, the post-shock $T_{EG}$, $T_{1T}$, $T_{tr,2T}$, $T_{tr,3T}$, and $T_{tr,3T,slip}$ increase due to shock relations. Then, they rise to reach a maximum after passing through the *RCW*, corresponding to the previously mentioned high-temperature center (see Fig. 26(a)); (b) $T_{ve,2T}$, $T_{v,3T}$, $T_{v,3T,slip}$, $T_{e,3T}$, and $T_{e,3T_{slip}}$ also increase after the shock, but are markedly lower than the $T_{tr}$, demonstrating significant non-equilibrium. (c) All $\gamma$ generally exhibit a decreasing trend. Their variation patterns also correspond to the patterns of respective temperatures (refer to the discussion of the distributions along *FL1* for details).

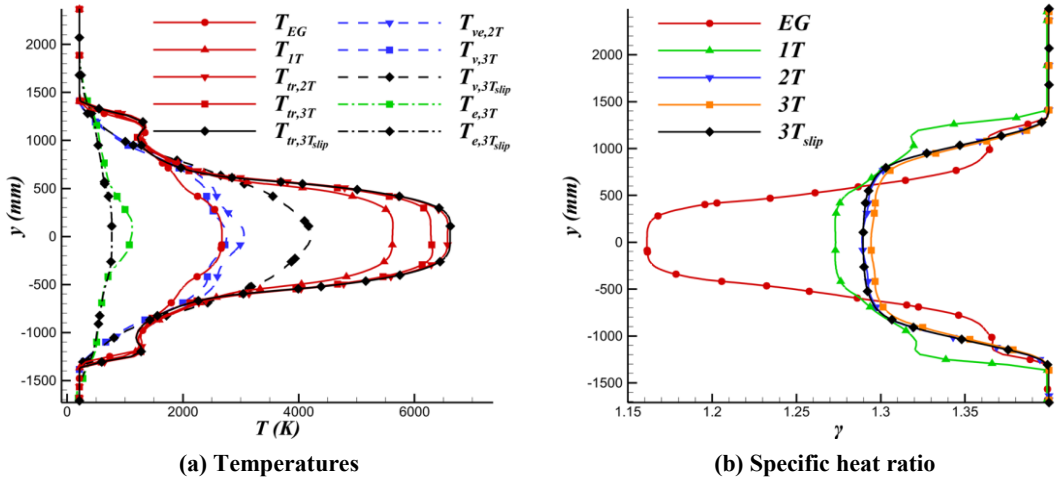

(a) Temperatures  (b) Specific heat ratio

**Fig. 29** Distributions of temperatures and $\gamma$ along *FL3* versus $y$ by different models at $\alpha = 0°$

(b) Species mass fraction

Given the similarity of $Y_i$ distributions among different models, Fig. 30 presents the distributions of $Y_O$, $Y_N$, $Y_{NO}$, and $Y_{NO^+}$ on the horizontal plane ($y = 0$) at $\alpha = 0°$, using the *2T*-M as an example. It can be observed that species generated in the upstream boundary layer are transported to the base region via flow separation and vortices, accumulating near the base centerline and forming a maximum structure. In purpose of a comparative analysis, using NO as an example, Fig. 31 further presents the difference between the *2T*- and *3T*-M results relative to the *1T*-M, as well as a comparison for the *3T*-M with and without slip conditions. The comparison reveals that:

(a) The $Y_{NO}$ at the base satisfy $Y_{NO,1T} > Y_{NO,2/3T}$, with the latter being generally smaller. This is primarily because the $T_c$ in Eq. (6) is relatively lower in the multi-temperature models; (b) The $Y_{NO,3T}$ is distributed relatively uniformly over the base wall and downstream, while the $Y_{NO,3T_{slip}}$ is very small near the base wall. But along the downstream centerline, the higher $T_{tr,3T_{slip}}$ and $T_{v,3T_{slip}}$ (Fig. 28(a)) leads to a higher $T_c$, which results in $Y_{NO,3T} < Y_{NO,3T_{slip}}$.

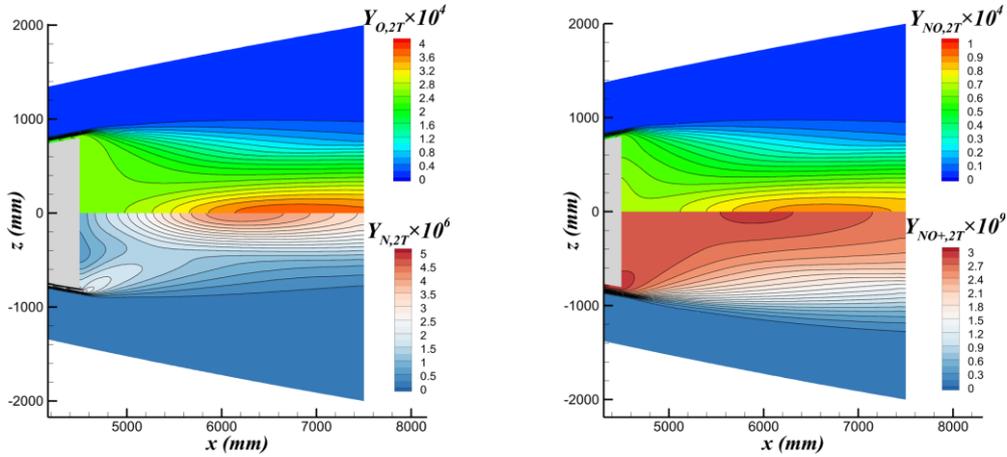

(a) $Y_{O,2T}$ (up) & $Y_{N,2T}$ (down)  (b) $Y_{NO,2T}$ (up) & $Y_{NO^+,2T}$ (down)

Fig. 30 Distributions of $Y_i$ on the horizontal plane ($y = 0$) at the base by the *2T*-M at $\alpha = 0°$

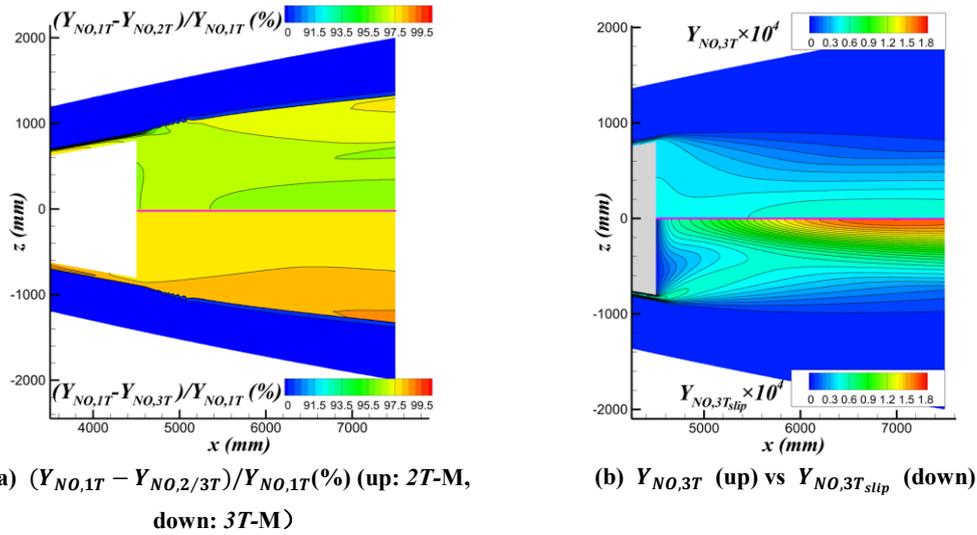

(a) $(Y_{NO,1T} - Y_{NO,2/3T})/Y_{NO,1T}$(%) (up: *2T*-M, down: *3T*-M)  (b) $Y_{NO,3T}$ (up) vs $Y_{NO,3T_{slip}}$ (down)

Fig. 31 Differences of $Y_{NO}$ distribution on the horizontal plane ($y = 0$) at the base by different models at $\alpha = 0°$

To further analyze the differences in $Y_i$ among different models, Fig. 32 presents a distribution comparisons of $Y_N$, $Y_{NO}$, and $Y_{NO^+}$ along *FL2* and *FL3* at $\alpha = 0°$, as predicted by the various models, where an inset in Fig. 32(b) displays the $Y_i$ distributions on a logarithmic scale to magnify the small differences. It can be observed that along *FL2*, as $x$ increases, the three components under all models exhibit relatively gentle variations on the logarithmic scale, while the results from the *3T$_{slip}$*-M show a distinct decrease near the wall; along *FL3*, they are concentrated between the two shock waves and reach their maxima at $y = 0$, which is consistent with the maximum structure shown in Fig. 30. It is worth noting that the $Y_i$ on two *FL*s indicate that, apart from $Y_{N,1T}$ being relatively large, the other $Y_i$ of all models are comparatively small.

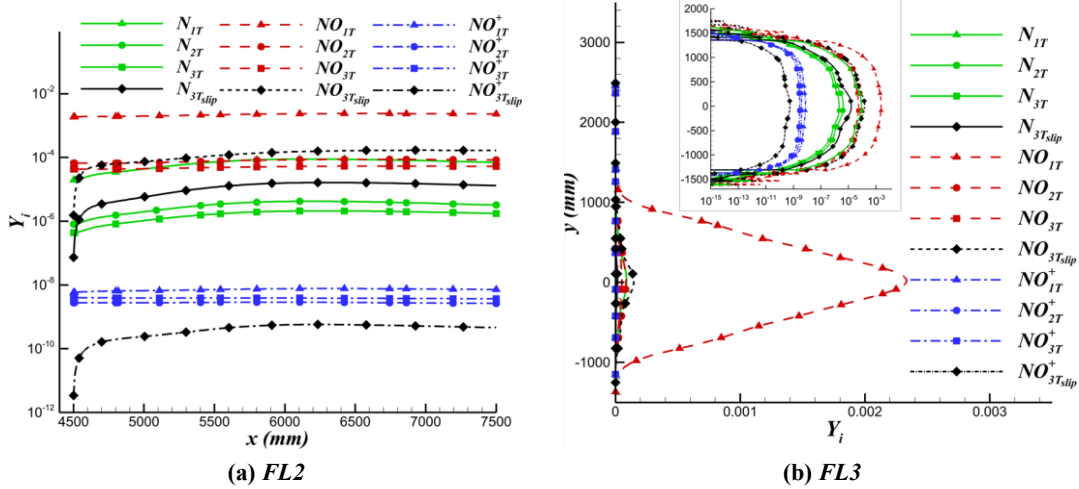

(a) *FL2*  (b) *FL3*

Fig. 32 Distributions of $Y_N$, $Y_{NO}$, and $Y_{NO^+}$ along feature lines at the base by different models at $\alpha = 0°$

(2) Case at $\alpha=8°$

(a) Temperatures and $\gamma$

Given the distribution similarity of $T$ & $T_{tr}$ as well as $T_{ve}$ & $T_v$ across the different models, $T_{tr,2T}$, $T_{ve,2T}$, and $T_{e,3T}$ are chosen as representatives with presents their distributions in *SMP* shown in Fig. 33. Considering that the cone flow is no longer axisymmetric, Fig. 34 further provides the corresponding distribution contours on the streamwise cross-sections at the cone base, as well as three-dimensional iso-surfaces. The latter respectively correspond to $T_{tr,2T}$ = 2000 & 6500 K, $T_{ve,2T}$ = 2100 & 3800 K, $T_{e,3T}$ = 600 & 1100 K, and $T_{e,3T_{slip}}$ = 600 & 800 K. From Fig. 34 (as well as Fig. 33), the following observations can be made: $T_{tr,2T}$ exhibits two maximum structures at the base which are symmetric to the *SMP*, appearing on the leeward side at the presence of AOA; $T_{ve,2T}$ shows a single strip-like structure with high intensity at the base, drawn from the boundary layer on the windward of the cone; the high-temperature region of $T_{e,3T}$ is concentrated near the cone base, and the structures with and without slip are similar.

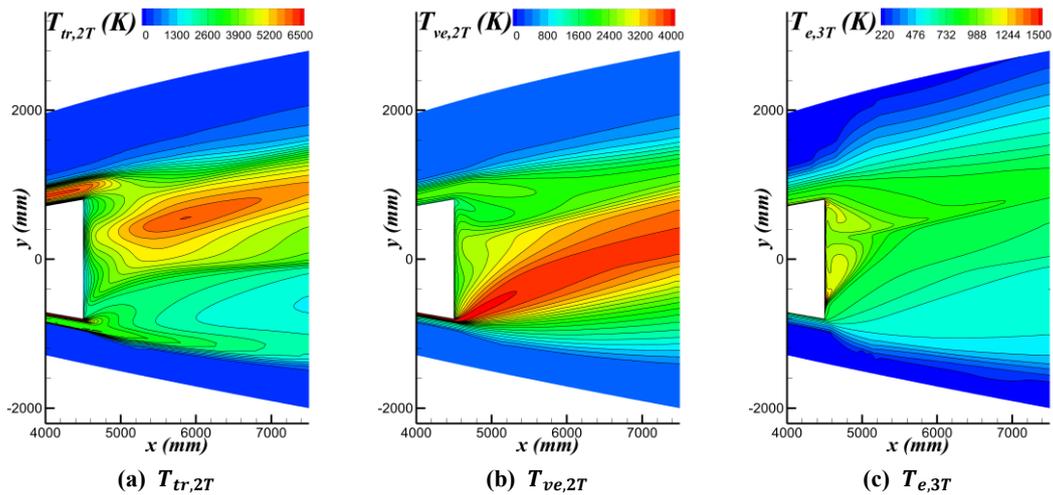

(a) $T_{tr,2T}$  (b) $T_{ve,2T}$  (c) $T_{e,3T}$

Fig. 33 Temperature distributions on the cone base *SMP* by the *2T*- and *3T*-M at $\alpha = 8°$

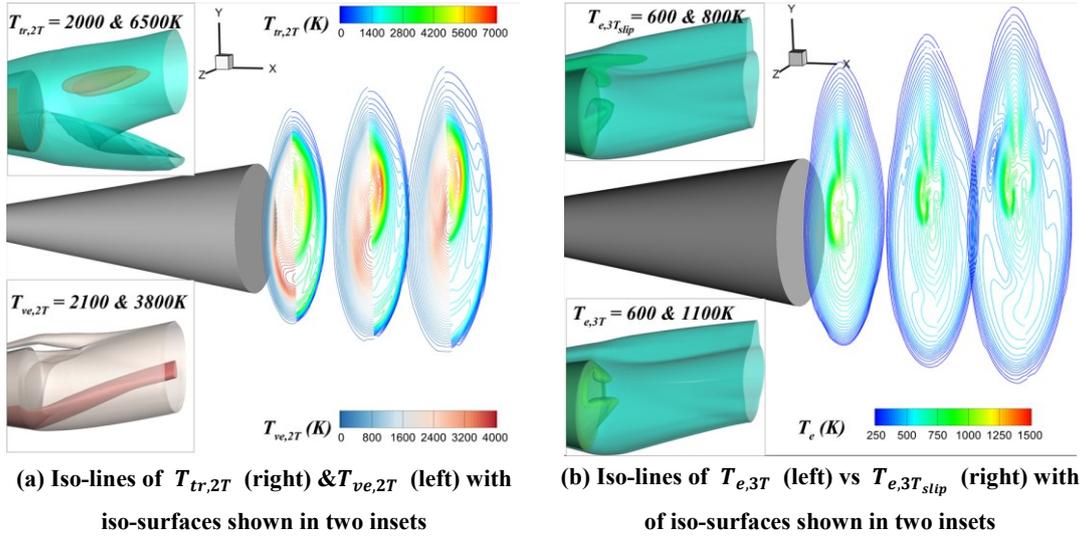

(a) Iso-lines of $T_{tr,2T}$ (right) & $T_{ve,2T}$ (left) with iso-surfaces shown in two insets

(b) Iso-lines of $T_{e,3T}$ (left) vs $T_{e,3T_{slip}}$ (right) with of iso-surfaces shown in two insets

Fig. 34 Temperatures contours on the streamwise cross-sections and iso-surfaces at the base by different models at $\alpha = 8°$

To quantify the differences of $T$ and $T_{tr}$ distributions at the base among the different models, we first present the magnitudes and locations of the temperature extrema ($T_{EXT}$) for $T$ and $T_{tr}$ at the base flow, as shown in Fig. 35. Notably, the $x$ and $y$ coordinates of the extremum points are not displayed in the figure due to their minimal variation across models. Given the extremum points are symmetric about $z = 0$, their $z$-coordinates are presented using $|z_{T_{EXT}}|$. The results indicate that compared to the 1T-M, the EGM yields smaller values for both $T_{EXT}$ and $|z_{T_{EXT}}|$. Among the different temperature models, the values of $|z_{T_{EXT}}|$ are quite similar, while the extrema magnitudes satisfy $T_{EXT,1T} < T_{tr_{EXT},3T} < T_{tr_{EXT},2T}$. For the $3T_{slip}$-M, the $T_{EXT}$ is greater than that of the non-slip case, while the $|z_{T_{EXT}}|$ values are approximately equal.

Secondly, focusing on the $T_{ve,2T}$ field in the base flow from Fig. 34(a), we extract the magnitudes and coordinates of the temperature extrema on a series of streamwise sections, and the same procedure is also imposed to the $3T$- and $3T_{slip}$-M field. Consequently, Fig. 36 presents the distributions of $T_{ve/v_{EXT}}$ extremum and their $y$-coordinates ($y_{T_{ve/v_{EXT}}}$) with $x$. It can be observed that as $x$ increases, the extrema by all models shift toward the leeward side, and $T_{ve/v_{EXT}}$ decreases gradually. At each streamwise section, the extremum positions of the 2T- and 3T-M are nearly identical, with the magnitudes satisfy $T_{v_{EXT},3T} < T_{ve_{EXT},2T}$. For the $3T_{slip}$-M, both $T_{v_{EXT}}$ and $y_{T_{v_{EXT}}}$ are larger than those of the non-slip counterpart. The underlying causes for the discrepancies have been analyzed in preceding sections and are therefore not reiterated here.

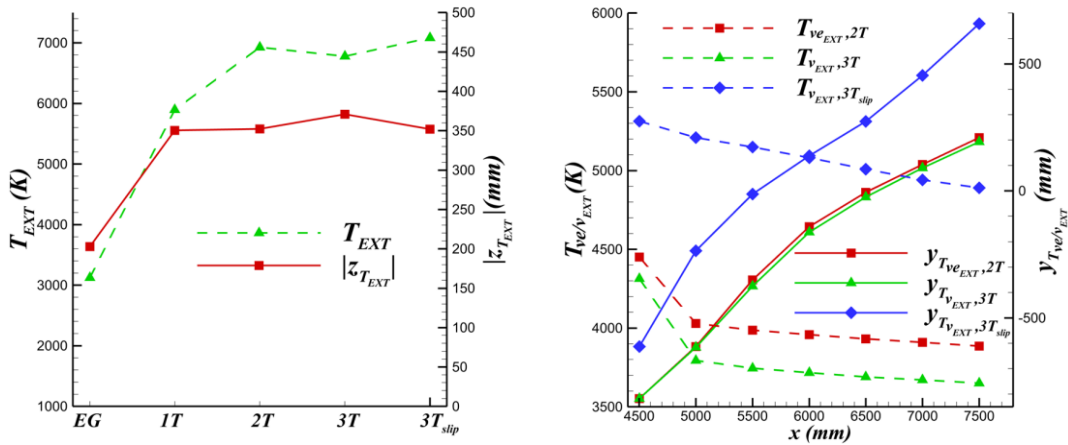

Fig. 35 Magnitudes and coordinates of $T$ and $T_{tr}$ extrema at the base by different models at $\alpha = 8°$

Fig. 36 Distributions of magnitude and $y$ of $T_{ve/v}$ extrema with $x$ by different models at $\alpha = 8°$

To quantitatively analyze the influence of different models on the cone base flow field at $\alpha = 8°$, Fig. 37 and Fig. 38 present distribution comparisons of various temperatures along $FL2$ and $FL3$, respectively. Since the $\gamma$ at both lines distributes overall similar to those at $\alpha = 0°$ (except for $\gamma_{EG}$ along $FL2$, which shows an initial decrease followed by an increase), they are not shown here but provided in the Appendix for reference. The results in Fig. 37 show that along $FL2$, as $x$ increases: (a) $T_{EG}$, $T_{1T}$, $T_{tr,2T}$, and $T_{tr,3T}$ increase rapidly near the base, reach a maximum at approximately $x \approx 5350$ mm, and then decrease slowly. The maximum temperatures among the different models satisfy $T_{EG} < T_{1T} < T_{tr,3T} < T_{tr,2T}$, with $T_{EG}$ being significantly lower than the latter three, which is consistent with the trend observed at $\alpha = 0°$; (b) The distributions of $T_{ve,2T}$ and $T_{v,3T}$ exhibit rapid initial growth followed by a slower increase, reaching their maxima at approximately $x \approx 6500$ mm. Their magnitudes satisfy $T_{ve,2T} > T_{v,3T}$, in agreement with the results in Fig. 21(d). Furthermore, both $T_{ve,2T}$ and $T_{v,3T}$ are lower than the corresponding $T_{tr}$, demonstrating pronounced non-equilibrium characteristics. The distribution of $T_e$ near the base is constrained by the boundary conditions, while it remains essentially constant further downstream; (c) Regarding the effect of slip flow, the distribution trend of $T_{tr,3T_{slip}}$ is similar to that of the non-slip result. It is noteworthy that, influenced by the energy jump boundary condition (see Eq. (14)), $T_{v,3T_{slip}}$ near the base is not only significantly higher than $T_{v,3T}$ under the isothermal wall condition but also exceeds all $T_{tr}$ in that region. $T_{e,3T_{slip}}$ essentially coincides with $T_{e,3T}$ in the downstream region but is relatively smaller in the vicinity of the base wall.

The results in Fig. 38 show that along $FL3$: (a) As $y$ increases, the freestream passes through the strong windward shock. According to the shock relations, the post-shock values of $T_{EG}$, $T_{1T}$, $T_{tr,2T}$, $T_{tr,3T}$, and $T_{tr,3T,slip}$ increase sharply. Then they decrease due to the expansion wave present at the outer edge of the cone windward side, and increase again through recompression to reach a maximum, which is consistent with the maximum structure of $T_{tr}$ shown in Fig. 33. Afterwards, as $y$ continues to increase, the distribution passes through the leeward expansion wave and shock, causing the temperature to decrease back to the freestream level; (b) $T_{ve,2T}$, $T_{v,3T}$, $T_{v,3T_{slip}}$, $T_{e,3T}$, and $T_{e,3T_{slip}}$ also exhibit growth between the upper and lower shock waves, but their distributions are distinctly different from those of $T_{tr}$, thereby revealing significant non-equilibrium characteristics. Furthermore, influenced by the energy jump boundary condition, $T_{v,3T_{slip}}$ near the base is also significantly higher than $T_{v,3T}$ under the isothermal wall condition.

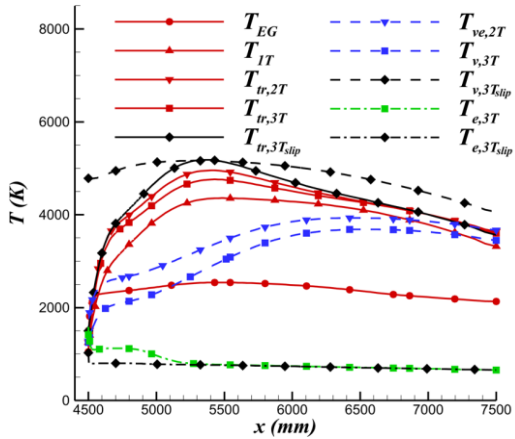
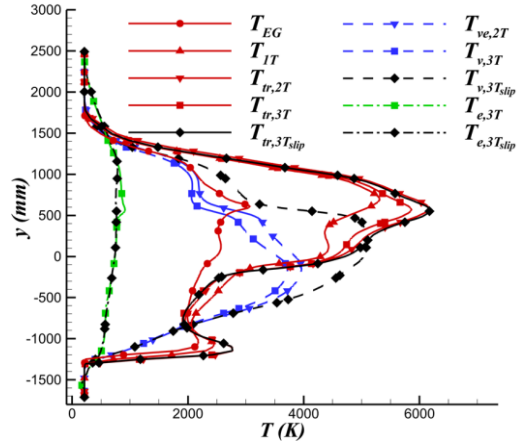

Fig. 37 Distributions of temperatures along $FL2$ at

Fig. 38 Distributions of temperatures along $FL3$ at cone base versus $x$ by different models at $\alpha = 8°$    cone base versus $y$ by different models at $\alpha = 8°$

(b) Species mass fraction

Since the structures of $Y_i$ are similar across the different temperature models, Fig. 39 presents the results of the *2T*-M as an example, namely, the contour distributions on streamwise sections at the cone base and their corresponding iso-surfaces for $Y_i$ at $\alpha = 8°$. The iso-surfaces correspond to $Y_{O,2T} \times 10^3 = 5.0$ & $9.0$, $Y_{N,2T} \times 10^3 = 0.3$ & $0.7$, $Y_{NO,2T} \times 10^3 = 3.0$ & $6.0$, and $Y_{NO^+,2T} \times 10^7 = 0.1$ & $0.7$. These iso-surfaces heuristically illustrate the origin of the aforementioned species at the base—they primarily originate from the transport of fluid drawn from the boundary layer on the windward side of the cone and are concentrated in corresponding structures, although their magnitudes are generally relatively small.

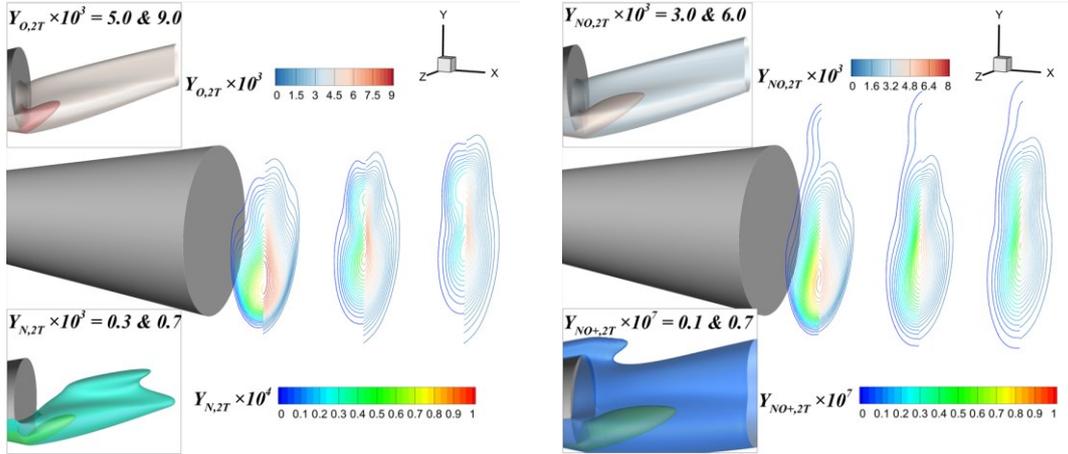

(a) Iso-lines of $Y_{O,2T}$ (right) & $Y_{N,2T}$ (left) with iso-surfaces shown in two insets

(b) Iso-lines of $Y_{NO,2T}$ (right) & $Y_{NO^+,2T}$ (left) with iso-surfaces shown in two insets

Fig. 39 Contours of $Y_i$ on streamwise sections and iso-surfaces at cone base by the *2T*-M at $\alpha = 8°$

To further quantitatively analyze the differences in $Y_i$ among the different models, Fig. 40 presents a comparison of the distributions of $Y_N$, $Y_{NO}$, and $Y_{NO^+}$ along *FL2* and *FL3* at $\alpha = 8°$, where a subplot in Fig. 40(b) displays the distributions on a logarithmic scale to reveal the features of the relatively small $Y_i$ values. It can be observed that the distribution trends of the three species along the two *FL*s under the different temperature models are similar to those at $\alpha = 0°$. However, due to the stronger windward shock wave under AOA, the $Y_i$ of productions drawn from the boundary layer on the same side is relatively larger.

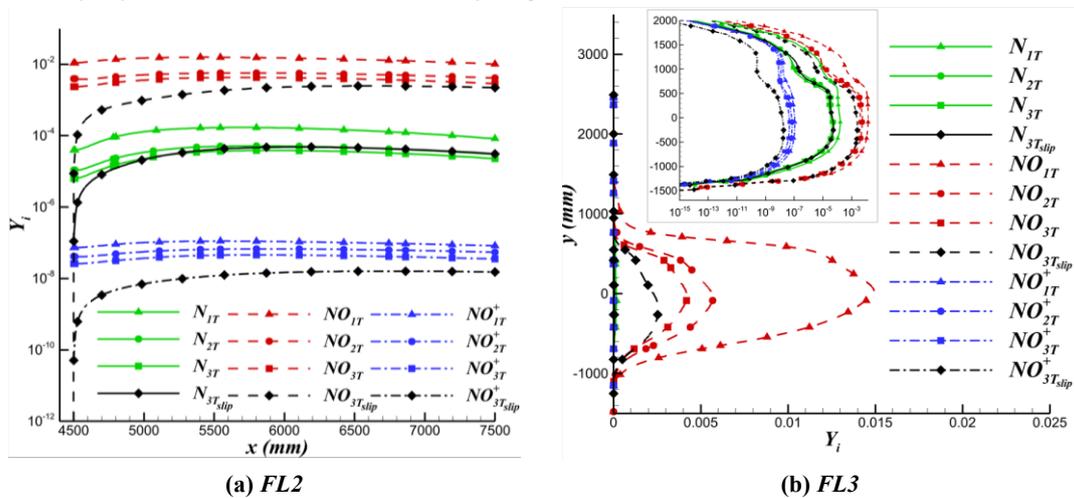

(a) *FL2*      (b) *FL3*

Fig. 40 Distributions of $Y_N$, $Y_{NO}$, and $Y_{NO^+}$ along *FL*s at cone base by different models at $\alpha = 8°$

# 4 Conclusion

This paper investigates the equilibrium and non-equilibrium flow characteristics predicted by different models and boundary conditions for a cone configuration under $M_\infty$ =27 and $H$ = 72 km. Numerical simulations based on the Navier-Stokes equations were conducted employing thermochemical equilibrium and non-equilibrium models (*1T*-, *2T*-, *3T*-, and *3T$_{slip}$*-M). Through comprehensive analysis of the flow structure, aerodynamic characteristics, and variable distributions, the following conclusions are drawn:

(1) Flow structures. At $\alpha$ = 0°, the conical flow structure includes a bow shock at the nose, a conical shock along the body, *EW* at the base, *RCW*, and two symmetric vortices. At $\alpha$ = 8°, similar shock structures are present where the windward shock is stronger and the shock layer is thinner, and the base vortices exhibits asymmetrically with a larger size in the lower part. The surface flow topology at the base features a primary node and two primary saddle points, with alternating secondary saddle points and nodes appearing along the separation streamlines. The overall flow structures predicted by the different models are similar. However, discrepancies exist in the locations of structures such as the high-temperature region at the base and the pressure center. Significant differences are observed in the near-wall distributions of $T_{v,3T}$ and $Y_i$ between the slip and non-slip boundary conditions.

(2) Aerodynamic characteristics. The $C_D$ predicted by different models vary due to differences in the $C_{D,f}$. The *EGM* yields the highest $C_D$, while the *3T*-M gives the lowest. Notably, the relative differences between the results of the *1T*- and *2T*-M compared to the *3T*-M exceed 5% at both AOAs, indicating that the physical model can significantly influence the prediction of aerodynamic characteristics. The distributions of $P_w$, $C_f$, and $Q_w$ predicted by the different models follow similar trends. Specifically, the *EGM* predicts higher $Q_w$ at the base at both AOAs, as well as higher $C_f$ and $Q_w$ in body section at $\alpha$ = 8°, compared to the non-equilibrium models. The $Q_w$ and $C_f$ of the *3T*-M are generally lower than those of the *1T*- and *2T*-M at both AOAs, while the $P_w$ show minimal variation across models. Regarding the effect of slipping, the *3T$_{slip}$*-M yields a $C_D$ close to the non-slip result, whereas $C_f$ and $Q_w$ of the former are slightly lower overall.

(3) Equilibrium and non-equilibrium characteristics. The discrepancies among different models are primarily manifested within the shock layer and the wake flow field, where the multi-temperature models exhibit distinct non-equilibrium features. Compared to the *1T*-M, the *EGM* exhibits a generally lower $\gamma$ and lower overall temperatures, with the maximum difference at the base reaching up to 3000 K. The $T_{tr}$ from multi-temperature models is higher and $T_{ve/v}$ is lower than the $T$ from the *1T*-M, meanwhile the $T_{e,3T}$ is even lower. The differences in specific heat capacity are predominantly determined by temperature, and therefore the multi-temperature models yield a higher $\gamma$ due to their lower $T_{ve/v}$ and consequently lower specific heat capacity. Furthermore, the higher $T_c$ from the *1T*-M leads to larger $Y_i$ compared to the multi-temperature models. For the *3T$_{slip}$*-M results, the jumps of temperature and energy at the wall leads to overall higher $T_{tr}$ and $T_v$ compared to the non-slip case. This is particularly evident along the centerline of the base, where $T_v$ is significantly higher than that from other models, reaching levels comparable to the $T$ and $T_v$ of *1T*-, *2T*-, *3T*-M at $\alpha$ = 8°.

## Appendix

Fig. 41 first shows the distributions of temperatures, $\gamma$, $u_t$, and $Y_i$ along the *FL1* on both the windward and leeward sides at $\alpha=8°$ by different models. Next, Fig. 42 provides distribution comparisons of $\gamma$ along *FL2* and *FL3* at $\alpha=8°$.

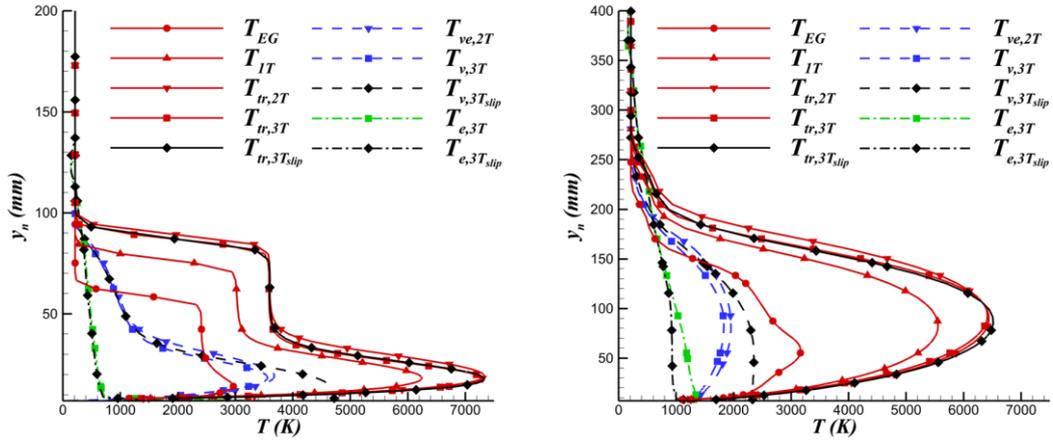

(a) Temperatures (left: windward, right: leeward)

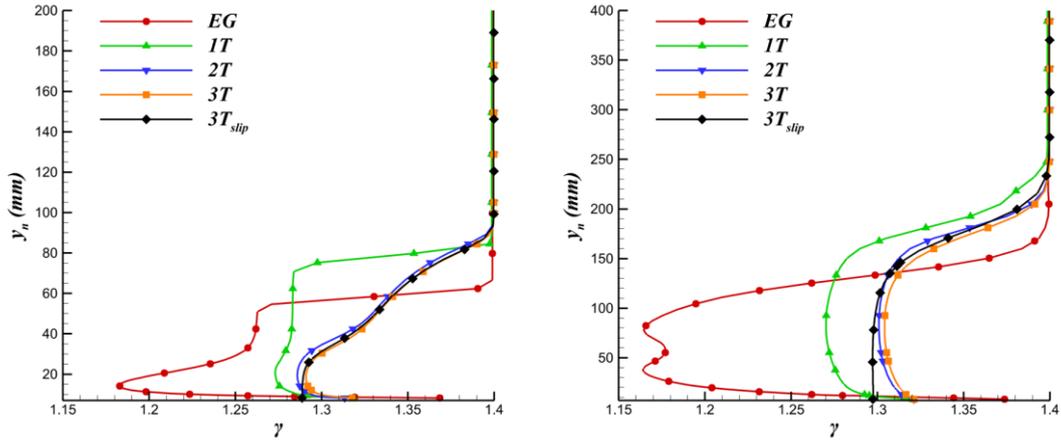

(b) Specific heat ratio (left: windward, right: leeward)

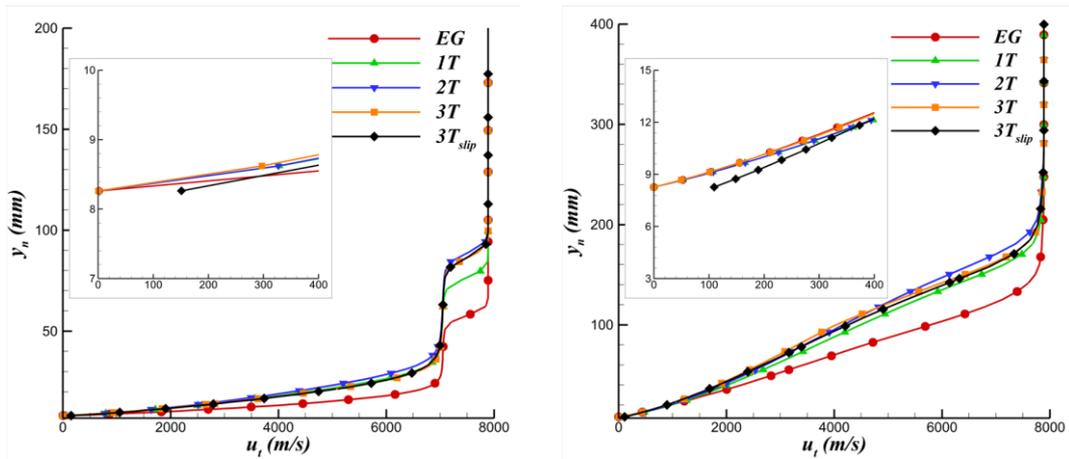

**(c) Velocity component parallel to the wall (left: windward, right: leeward)**

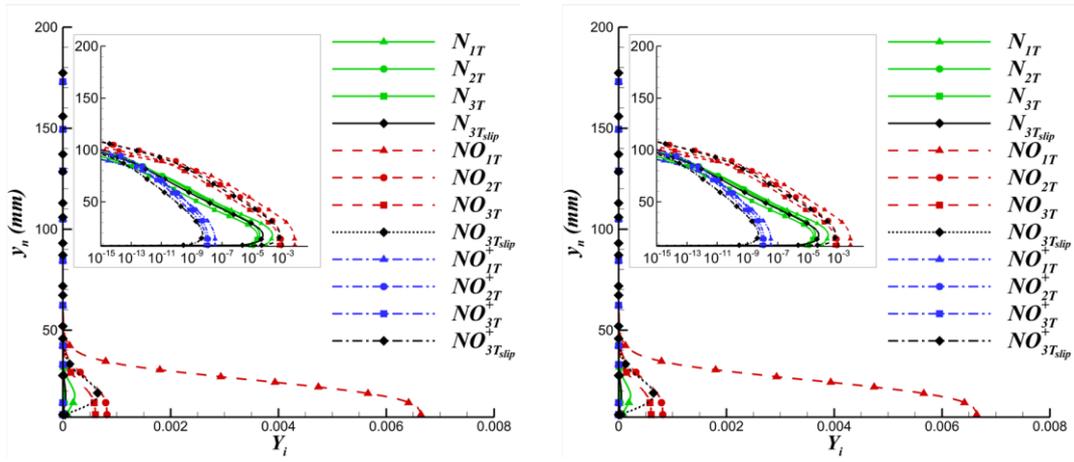

**(d) Species mass fraction (left: windward, right: leeward)**

**Fig. 41 Distributions of quantities along *FL1* by different models at $\alpha = 8°$**

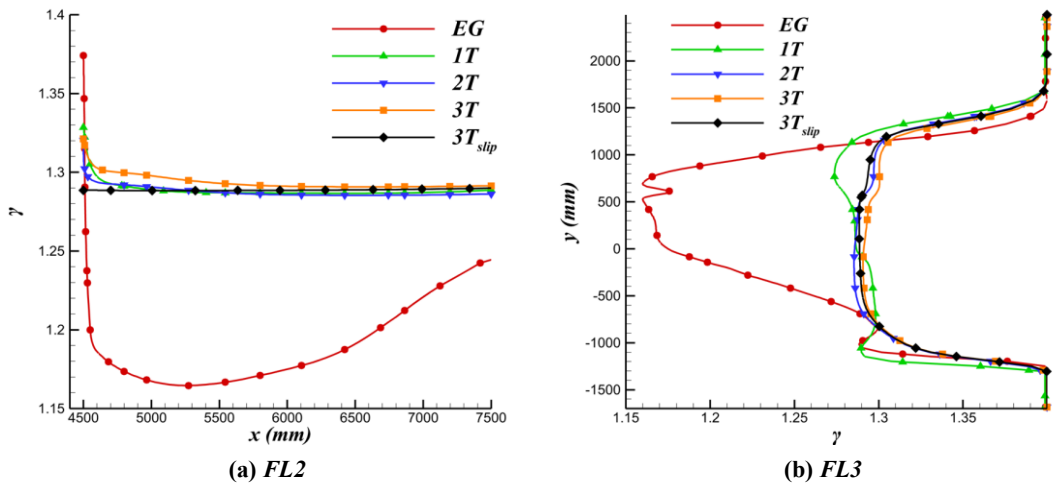

**(a) *FL2***  **(b) *FL3***

**Fig. 42 Distributions of the specific heat ratio along *FL*s at cone base by different models at $\alpha = 8°$**